\documentclass[11pt]{article}

\usepackage[final]{acl}

\usepackage{times}
\usepackage{latexsym}

\usepackage[T1]{fontenc}

\usepackage[utf8]{inputenc}

\usepackage{microtype}

\usepackage{inconsolata}

\usepackage{graphicx}

\usepackage{algorithmic}
\usepackage{algorithm}
\usepackage{microtype}
\usepackage{graphicx}
\usepackage{subcaption}
\usepackage{booktabs} 
\usepackage{twemojis}
\usepackage{xspace}
\usepackage{makecell}
\usepackage{longtable}
\usepackage{amsmath}
\usepackage{amssymb}
\usepackage{mathtools}
\usepackage{amsthm}
\usepackage{enumitem}
\usepackage{booktabs,multirow,diagbox,enumitem}
\usepackage[most]{tcolorbox}
\usepackage[capitalize,noabbrev]{cleveref}
\theoremstyle{plain}

\theoremstyle{definition}

\theoremstyle{remark}

\usepackage[table]{xcolor} 
\usepackage{tabularx}
\newcommand{\cmark}{\ding{51}} 
\newcommand{\xmark}{\ding{55}} 
\usepackage{pifont}
\newcommand{\vpara}[1]{\vspace{0.05in}\noindent\textbf{#1 }}
\newcommand{\HIDE}[1]{} 

\newcommand{\ours}{RAT\xspace}
\newcommand{\oursbench}{RATBench\xspace}

\title{RAT: RunAnyThing via Fully Automated Environment Configuration}

\author{
\textbf{Renhong Huang\textsuperscript{1,2}},
\textbf{Dongdong Hua\textsuperscript{1}},
\textbf{Yifei Sun\textsuperscript{1}},
\textbf{Sitao Ding\textsuperscript{1}},
\\
\textbf{Hanyang Yuan\textsuperscript{1}},
\textbf{Daixin Wang\textsuperscript{2}},
\textbf{Yang Yang\textsuperscript{1}\thanks{~~Corresponding author.}}
\\
\textsuperscript{1} Zhejiang University,
\textsuperscript{2} Ant Group.
\\
 \texttt{\{renh2,ddhua,yifeisun,martinding,yuanhanyang,yangya\}@zju.edu.cn},\\
  \texttt{\{daixin.wdx\}@antgroup.com}.
}

\begin{document}
\maketitle

\begin{abstract}
Automating repository-level software engineering tasks is a foundational challenge for autonomous code agents, largely due to the difficulty of configuring executable environments. However, manual configuration remains a labor-intensive bottleneck, necessitating a transition toward fully automated environment configuration. Existing approaches often rely on pre-defined artifacts or are restricted to specific programming languages, limiting their applicability to diverse real-world repositories. In this paper, we first propose \textbf{\ours} (RunAnyThing), a modular and extensible agent framework for fully automated configuration across programming languages on arbitrary repositories. \ours adopts a multi-stage pipeline that integrates language-aware abstraction, image initialization, specialized configuration toolset, and robust sandbox. Furthermore, to enable rigorous evaluation, we propose \textbf{\oursbench}, a benchmark reflects the comprehensive coverage of real-world repositories. Extensive experiments demonstrate that \ours achieves state-of-the-art performance, improving Environment Setup Success Rate (ESSR) by an average of 36.1\% over strong baselines.
\end{abstract}

\section{Introduction}
The evolution of Large Language Models (LLMs) has shifted the frontier of autonomous programming from simple snippet generation~\cite{zhu2022multilingual,bappon2024autogenics,coignion2024performance} to complex, repository-level engineering \cite{zhang2023repocoder,jimenez2023swe,shrivastava2023repository,wu2024repoformer}. However, unlike code snippets, repository-level tasks demand strict adherence to intricate inter-dependencies and environment-specific configurations. Without an executable environment, even logically correct code remains unverifiable and functionally invalid. Consequently, environment configuration has emerged as a key bottleneck in autonomous agents~\cite{hu2025repo2run}.

Moreover, environment configuration is not merely a matter of software convenience, but a fundamental requirement for the development of code LLMs~\cite{li2023starcoder,le2022coderl,luo2023wizardcoder}. Generally, automated environment configuration plays three critical roles: (1) \emph{Scalable Data Synthesis}: It enables scalable benchmark construction by transforming static repositories into verifiable datasets, rather than relying on manually curated benchmarks such as SWE-bench~\cite{jimenez2023swe}. In addition, it helps synthesize accurate execution traces essential for LLM training~\cite{da2025agent}. (2) \emph{Execution-based Reinforcement}: It supports functional feedback loops, allowing reward models to move beyond static or heuristic signals~\cite{le2022coderl} toward true executability. (3) \emph{System Reliability}: It ensures deployment reproducibility, overcoming the inherent brittleness of CI/CD integration. Overall, automated environment configuration is key to transforming code agents from symbolic generators into reliable autonomous systems.

Although several pioneering efforts have attempted to automate environment setup, they often rely on strong assumptions that limit their generalization to real-world repositories. For instance, INSTALLAMATIC~\cite{milliken2025beyond} and EXECUTIONAGENT~\cite{bouzenia2025you} rely on pre-existing artifacts, such as curated Dockerfiles, installation scripts, or CI logs. While Repo2Run~\cite{hu2025repo2run} introduces dual-environment architecture to decouple configuration from monitoring, it still operates within a rigid framework that lacks the flexibility to handle diverse, uncurated repositories. Overall, these methods are ill-suited for scaling to thousands of real-world repositories due to their dependence on specific prior knowledge and limited language support. 

To overcome the limitations of existing approaches, we introduce \ours (RunAnyThing), a modular and extensible agent framework for fully automated environment configuration across programming languages. \ours employs an LLM-driven multi-stage pipeline starting with ImageRetriever, which semantically analyzes repositories to select optimal base images and reduce configuration overhead. Different configuration mode enable handling of complex or previously unseen repositories, while configuration toolset and expertise accumulation resolve configuration ambiguities and retain knowledge across sessions. Together, these components provide scalable, adaptive, and robust environment configuration for diverse repositories.

Furthermore, to rigorously evaluate environment configuration methods under realistic repository settings, we introduce \oursbench, a large-scale multilingual benchmark comprising over 2,500 GitHub repositories. Unlike existing datasets that are limited in language coverage or biased toward trivial projects, \oursbench is constructed via stratified sampling to capture real-world diversity in project distribution, programming languages, and availability, and is validated through a rigorous executability-driven pipeline. Extensive experiments on \oursbench demonstrate that \ours achieves a state-of-the-art Environment Setup Success Rate (ESSR) and exhibits environment configuration capabilities that surpass human experts. 
\section{Preliminary}\label{sec:pre}
\vpara{Configuration Artifacts.}We refer to files that specify environment setup as \emph{configuration artifacts}. Typical artifacts include Dockerfiles, CI pipelines, build manifests (e.g., \texttt{package.json}, \texttt{pom.xml}, \texttt{Cargo.toml}), lockfiles (e.g., \texttt{poetry.lock}), and explicit test scripts. When present, these artifacts partially or fully define the required runtime environment (e.g., language and library versions) and the verification sequence. When artifacts are absent or incomplete, required environment and verification sequence must be inferred from documentation (e.g., \texttt{README}) and repository structure.

\vpara{Environment Configuration.} Let $\mathcal{R}$ denote a repository comprising source code and metadata, and let $\mathcal{E}(\mathcal{R})$ denote the set of feasible containerized environments compatible with $\mathcal{R}$. The goal of environment configuration is to construct an environment $e \in \mathcal{E}(\mathcal{R})$ along with a verification sequence $\pi(\mathcal{R})$, where $\pi(\mathcal{R})$ specifies the repository-dependent execution procedure (e.g., tests or build commands).

A configuration is considered successful if the execution of $\pi(\mathcal{R})$ within $e$ terminates without error. The output of the task is either a runnable container image or Dockerfile that deterministically builds such image.

\vpara{Execution Trace.} An environment configuration session induces an interaction trace $\tau(\mathcal{R})=((a_1,o_1),\ldots,(a_T,o_T))$ , where each action $a_t$ is a tool invocation and each observation $o_t$ is the resulting system feedback. The trace represents the agent’s interaction history with the environment and serves as the basis for subsequent decisions.
\section{Framework: \ours}
\begin{figure*}[!t]
    \centering
    \includegraphics[width=1.0\textwidth]{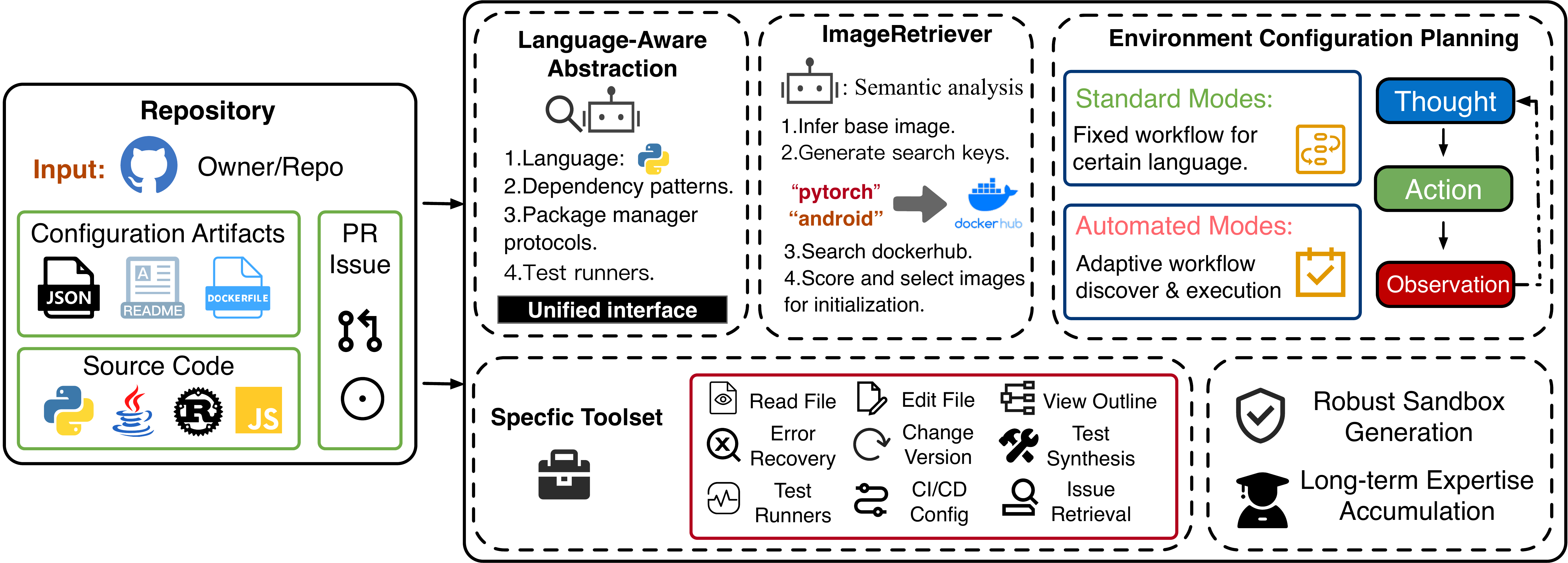}
    \caption{Overview of \ours (RunAnyThing) architecture. The framework consists of several core modules: (1) Language-Aware Abstraction, which isolates language-specific execution from general agent capabilities via a unified interface; (2) ImageRetriever, which analyzes repository semantics to select optimal base images, thereby reducing configuration overhead and improving success rates; (3) Configuration Modes, combining fixed workflow based setup with adaptive, repository-driven automation for flexible environment configuration; and (4) Configuration Toolset, which abstracts low-level terminal operations into high-level actions within a robust sandbox environment.}\label{fig:framework}
\end{figure*}

As shown in Figure~\ref{fig:framework}, we introduce \ours (\textbf{R}un\textbf{A}ny\textbf{T}hing), a modular and extensible agent framework designed for fully automated configuration of complex execution environments. The framework consists of a multi-stage pipeline within containerized sandboxes, including  language-aware abstraction, image initialization, specialized configuration toolset, and expertise accumulation. 

\vpara{Language-Aware Abstraction.}Repository environment configuration involves language-dependent components (e.g., dependency specifications, package managers, and testing protocols), while the overall automation workflow remains largely shared across repositories. To decouple language-specific logic from general agent capabilities, \ours introduces a modular abstraction layer that isolates language-specific heuristics into reusable components, improving extensibility and simplifying support for new programming languages.

Specifically, at the onset of configuration, \ours identifies the programming languages in a repository and encapsulates language-specific heuristics, including dependency patterns (e.g., \texttt{pom.xml} for Java and \texttt{Cargo.toml} for Rust), package manager protocols, and test runners, into a unified interface. This allows language-specific toolchains to be invoked through a shared execution protocol while preserving language-aware behavior. Thus, the modular design is naturally extensible, as we demonstrate with a real implementation where adding a new programming language requires only minimal engineering effort (see Appendix~\ref{app:example}). Furthermore, as repositories often contain multiple languages, \ours selects the dominant language based on code proportion as the primary execution target.

\vpara{ImageRetriever for Initialization.}Environment configuration can benefit from leveraging existing images, particularly for complex repositories that share common architectural patterns. Given a repository, \ours first establishes the Language-Aware Abstraction module. The ImageRetriever module then performs LLM-based semantic analysis over repository documentation and configuration artifacts to infer optimal execution requirements, including programming language versions, operating system variants, deep learning frameworks, and critical dependencies.

Based on the above analysis, it retrieves candidate base images from a predefined pool of standard images (e.g., \texttt{python:3.10}, \texttt{openjdk:17}). For more complex projects, it further generates search queries to retrieve specialized images from Docker Hub. An LLM-based scoring mechanism is then applied to select the most suitable initialization image. Overall, ImageRetriever improves environment initialization quality and reduces configuration overhead. Detailed discussion on the effectiveness of initialization is shown in Appendix~\ref{app:discussion}.

\vpara{Configuration Modes.} The agent’s execution planning is fundamental to the environment configuration process. \ours supports two configuration modes with different levels of flexibility:

\begin{itemize}[leftmargin=10pt, nosep] 
\item \emph{Standard Mode}: The structured mode follows a fixed workflow for environment configuration. The agent analyzes the repository structure and iteratively executes configuration commands via terminal interface to resolve dependencies in a deterministic, language-aware manner.

\item \emph{Automated Mode}: To accommodate real-world scenarios with unpredictable requirements, \ours introduces automated  mode. Instead of language-specific configuration pipelines, the agent interacts autonomously to discover repository-specific workflows, enabling adaptive command execution and flexible environment setup under diverse requirements.
\end{itemize}

These two modes are exposed as alternative execution strategies and can be selected according to repository characteristics and configuration requirements. In addition, we adopt the ReAct~\cite{yao2022react} framework for $\tau(\mathcal{R})$, which structures interaction as sequences of thoughts, actions (e.g., configuration commands), and observations. This paradigm is well-suited for environment configuration: the explicit ``thought'' process enables reasoning over the current configuration state, while real-time observations provide immediate feedback from the terminal, ensuring synchronization between the LLM and the evolving system state.

\vpara{Configuration Toolset.} To facilitate automated repository analysis and environment configuration,  we design a comprehensive toolset covering repository understanding, knowledge retrieval, environment setup, and validation. Unlike conventional approaches that rely on raw terminal commands, our tools are tightly integrated and tailored for configuration. Each tool abstracts low-level operations into a configuration-aware interface, which improves LLM context management through precise control of tool outputs and functionality. We provide representative examples in Figure~\ref{fig:framework}, while complete tool specifications and design principles are deferred to Appendix~\ref{app:tool_specifications} and Appendix~\ref{app:discussion}, respectively, due to space constraints.

\vpara{Robust Sandbox Generation.} Based on initialization, \ours~leverages a template-based Docker generation mechanism. This module constructs a tailored Dockerfile that automates the installation of the required runtime, configures localized mirrors for network connectivity, and injects the \ours~utility toolset into the container. To guarantee reliability, each environment undergoes pre-flight build validation before deploying the agent into the sandbox.

\vpara{Long Term Expertise Accumulation.} As the adage goes, ``Practice makes perfect''. Effective environment configuration is a knowledge-intensive process that scales with exposure to diverse repository structures. To formalize this, we introduce an automated mechanism for agents to synthesize expertise from historical execution trajectories. This accumulated experience is structured into a serialized schema (e.g., JSON), facilitating high-precision configuration. 

\section{Evaluation: \oursbench}\label{sec:bench}
\begin{table*}[!t]
    \centering
    \small
    \setlength{\tabcolsep}{6pt}
    \caption{\textbf{Benchmark comparison.} \textbf{\# Repos}: total number of repositories. \textbf{Langs.}: programming languages covered (P: Python, J: Java, K: Kotlin, R: Rust, JS/TS: JavaScript/TypeScript, G: Go). \textbf{Stratified}: whether repositories are sampled to balance repository size and popularity. \textbf{Auto-Collect}: whether repositories are mined from GitHub via automated collection. \textbf{Exec-Verified}: whether repository validity is assessed via executing builds/tests rather than static analysis. \textbf{Difficulty Levels}: whether explicit difficulty levels are provided. \cmark/\xmark~indicate presence or absence of the feature.}
    \renewcommand{\arraystretch}{1.25}
    \begin{tabularx}{\textwidth}{>{\raggedright\arraybackslash}Xcccccc}
    \toprule[1.2pt]
    \rowcolor{gray!15} \textbf{Benchmark} & \textbf{\# Repos} & \textbf{Langs.} & \textbf{Stratified} & \textbf{Auto-Collect} & \textbf{Exec-Verified} & \textbf{Difficulty Levels} \\ 
    \midrule
    \textbf{\oursbench} & 2,500+ & P, J, R, JS/TS, G & \cmark & \cmark & \cmark & \cmark \\ 
    \midrule
    EnvBench~\cite{eliseeva2025envbench} & 994 & P, J, K & \xmark & \cmark & \xmark & \xmark \\ 
    Repo2Run~\cite{hu2025repo2run} & 420 & P & \xmark & \cmark & \cmark & \xmark \\ 
    ExecutionAgent~\cite{bouzenia2025you} & 50 & 14 Langs & \xmark & \xmark & \cmark & \xmark \\ 
    Beyond Pip~\cite{milliken2025beyond} & 40 & P & \xmark & \xmark & \cmark & \xmark \\ 
    \bottomrule[1.2pt]
    \end{tabularx}
    \label{tab:benchmark_comparison}
\end{table*}

We next systematically evaluate the effectiveness of environment configuration methods. While several configuration benchmarks have been proposed, existing benchmarks are insufficient to fully evaluate a method's capacity for environment configuration. For instance, EnvBench~\cite{eliseeva2025envbench} relies on language-specific static metrics, such as missing import checks in Python or compilation checks in JVM, which often overlook complex runtime dependencies. Repo2Run~\cite{hu2025repo2run} focuses on the validity of generated Dockerfiles rather than the actual execution of tasks, while benchmarks like Beyond Pip~\cite{milliken2025beyond} rely on small-scale, manually curated samples, a labor-intensive process that inherently limits scalability and prevents comprehensive evaluation across diverse repositories.

To enable rigorous evaluation and overcome the above limitations, we introduce \oursbench, a large-scale benchmark comprising over 2,500 GitHub repositories. Unlike existing benchmarks that focus on limited languages or static settings, \oursbench is designed to reflect the complexity of real-world software repositories in terms of distribution, programming languages, and availability. Moreover, we develop a rigorous construction pipeline that ensures functional validity through automated collection and validation. A detailed comparison with existing benchmarks is provided in Table~\ref{tab:benchmark_comparison}.

\vpara{Diversity in Distribution.}
Existing benchmarks often suffer from bias toward trivial or popular projects. To mitigate this, we employed a two-dimensional stratified sampling strategy, spanning across multiple tiers of project size (ranging from lightweight utilities to large-scale systems) and popularity (spanning long-tail projects to top-tier repositories). This grid-based sampling ensures broad coverage of software complexity and prevents the evaluation from being dominated by simple or overly curated examples. As further evidenced in Appendix~\ref{app:exp}, this design enables effective coverage of repositories with diverse difficulty levels. Detailed distribution statistics are also provided in Appendix~\ref{app:bench_stats}.

\vpara{Diversity in Programming Languages.} 
\oursbench encompasses five widely used languages (Python, Java, Rust, JavaScript/TypeScript and Go) to capture diverse real-world environment failure modes. These languages span diverse interpreted and compiled toolchains, leverage different dependency managers, and present distinct configuration challenges. Specifically: (1) Python requires runtime verification as import graphs and optional native dependencies are often resolved only during execution, with failures frequently caused by missing system libraries. (2) Java exhibits build-lifecycle complexities (e.g., Maven or Gradle) and frequent dependency conflicts within multi-module projects. (3) Rust provides strong compiler guarantees but imposes strict toolchain and linker constraints, especially on target triples and native library linking. (4) JavaScript/TypeScript combines rapid runtime evolution, transpilation overhead, and native modules (e.g., \texttt{node-gyp}), making configurations highly sensitive to Application Binary Interface (ABI) versions and lockfile consistency. (5) Go features a comparatively streamlined build system, but environment failures still arise from module resolution, version constraints, build tags, cross-compilation settings, and native dependencies introduced via \texttt{cgo}.

\vpara{Diversity in Availability.}
To capture varying degrees of environment ambiguity, \oursbench categorizes repositories by the availability of (i) functional containerization artifacts and (ii) the inclusion of unit tests. These settings define the availability-based scenarios used in \S~\ref{subsec:metrics}, where fewer artifacts require stronger inference from repository structure and documentation.

\vpara{Automated Collection.}
We searched GitHub for repositories active within the past year, applying a minimum threshold of 10 stars to filter out obsolete or low-quality projects. To guarantee executability, we applied language-specific heuristics: repositories were required to contain standard build manifests (e.g., \texttt{pom.xml} for Java, \texttt{Cargo.toml} for Rust, \texttt{package.json} for Node.js) or explicit test directories (e.g., \texttt{tests/}, \texttt{test\_*.py}). For Python, repositories with Dockerfiles or CI/CD configurations were prioritized as they provide reliable environment ground truth.
\section{Experiments}\label{sec:exp}
In this section, we evaluate \ours and baseline methods on \oursbench under real-world environment configuration challenges. We further compare backbone models and evaluate \ours against human engineers. Additional results are provided in Appendix~\ref{app:exp}, including failure analysis, execution trajectory case studies, tool-call analysis, performance on other benchmarks, and other additional analyses. Code and dataset are available at \url{https://github.com/gemelom/RunAnyThing}.

\subsection{Evaluation Metrics.}\label{subsec:metrics}
\vpara{Environment Setup Success Rate (ESSR).}
We evaluate the efficacy of environment configuration using the Environment Setup Success Rate (ESSR), which measures the fraction of successfully passed unit tests within a configured environment. For a repository with $N$ unit tests, a naive definition would be: $\text{ESSR} = N_{\text{pass}}/{N}$, where $N_{\text{pass}}$ denotes the number of tests that pass successfully. However, to account for the fact that real-world repositories often contain pre-existing bugs or broken tests, we specifically refine this metric for Python repositories as $\text{ESSR} = N_{\text{pass}} / N_{\text{verified}}$ to account for potential noise or defects in real-world ground-truth artifacts. We report ESSR under three scenarios:
\begin{itemize}[leftmargin=15pt, nosep]
\item \textbf{S1 (Artifact-guided):} Repositories provided with unit tests and functional containerization artifacts. Here, $N_{\text{verified}}$ is the total number of existing unit tests, which serve as the gold-standard baseline.
\item \textbf{S2 (Artifact-free):} Repositories containing unit tests but lacking containerization scripts. To isolate failures caused by misconfiguration, $N_{\text{verified}}$ excludes tests that fail due to inherent code defects, even when executed in a manually verified environment.
\item \textbf{S3 (Test-deficient):} Repositories lacking both pre-existing tests and scripts. In this underspecified setting, we construct $N_{\text{verified}}$ by identifying runnable entry points or synthesizing smoke tests, defining success by the correct execution of these ad-hoc verification targets.
\end{itemize}

As for Java, Rust, and JS/TS repositories, we evaluate repositories with a deterministic build target. Success is defined by completing the corresponding build command without error (e.g., \texttt{mvn clean install} or \texttt{gradle clean build} for Java, \texttt{cargo build} for Rust, and \texttt{npm install} or \texttt{yarn install} for JS/TS). Detailed discussion on the rationale of the metric is provided in Appendix~\ref{app:discussion}.

\vpara{Efficiency Metrics.} In addition to assessing environment configuration effectiveness, we measure Latency (average execution time per repository) and Tokens (average token usage per repository) to quantify the practical computational overhead and deployment cost of each method.

\subsection{Baselines.} We evaluate our approach against five representative baselines, grouped into three categories: static \& prompt-based, software engineering agent and environment configuration agent. For \textbf{static \& prompt-based}, we compare with (1) pipreqs\footnote{Generate requirements.txt file for any project based on imports in \url{https://github.com/bndr/pipreqs}}, a traditional static analysis tool that generates dependency files by scanning source code imports; (2) Zero-shot LLM, which generates configuration scripts directly from README files without environment feedback; For \textbf{software engineering agent}, we compare with (3) SWE-agent~\cite{yang2024swe}, a general-purpose software engineering agent that handles repository-level tasks via interactive shell commands; For \textbf{environment configuration agent}, we compare with (4) Installamatic~\cite{milliken2025beyond}, a specialized agent for Python utilizing standardized installation contexts; (5) Repo2Run~\cite{hu2025repo2run}, a state-of-the-art agent that iteratively synthesizes Dockerfiles using an adaptive feedback loop and dual-environment execution. Besides, we further evaluate strong general-purpose coding agents (e.g., Claude Code) in Appendix~\ref{app:exp}.

\subsection{Experimental Results}\label{subsec:exp-res}
\vpara{Main Result.}
Table~\ref{tab:main-results-single-column} reports the ESSR across multiple programming languages. The results show that \ours obtains the highest ESSR among the evaluated methods in Table~\ref{tab:main-results-single-column}. On Python repositories, \ours achieves an ESSR of 63.2\%, significantly surpassing the traditional static analysis tool pipreqs. Moreover, compared to general-purpose SWE-agent, \ours yields an average improvement of 36.1\% across all evaluated programming languages. These results indicate that the environment configuration design of \ours is substantially more robust than general-purpose code agents when handling complex dependency structures. Furthermore, \ours consistently outperforms specialized environment configuration agents, underscoring the effectiveness and its advantage in configuring complex, multi-language environments in real-world repositories, without being limited to specific programming languages or setup scenarios.

Table~\ref{tab:rat-python-difficulty} compares \ours across three scenarios (S1–S3). Strong performance in S2 shows that \ours can effectively leverage project files even without containerization scripts. Meanwhile, the high ESSR in S3 indicates that \ours may autonomously infer entry points and generate effective smoke tests without relying on existing configurations. In contrast, Repo2Run suffers substantial performance degradation as configuration artifacts decrease, whereas \ours maintains strong performance, demonstrating robustness.

\begin{table}[!t] 
\centering
\footnotesize 
\caption{Environment setup success rate (ESSR, \%, higher is better) on \oursbench across various programming languages. \textbf{Bold}: best performance in each column. `/' denotes that the method is not applicable.}
\label{tab:main-results-single-column}
\setlength{\tabcolsep}{2.5pt}
\resizebox{\columnwidth}{!}{
\begin{tabular}{llccccc}
\toprule
\multicolumn{2}{c}{\textbf{Model Configuration}} & \multicolumn{5}{c}{\textbf{Programming Languages}} \\
\cmidrule(lr){1-2} \cmidrule(lr){3-7}
Framework & LLM & Python & Java & Rust & JS/TS & Go \\  
\midrule
\multicolumn{7}{c}{\textit{Static \& Prompt-based}} \\
\midrule
pipreqs & \cellcolor{gray!5} None  & 35.8 & / & / & / & / \\
\multirow{1}{*}{Zero-shot} 
  & \cellcolor{gray!10} DeepSeek-V3 & 15.2 & 0.0 & 0.0 & 7.3 & 0.0\\
\midrule
\multicolumn{7}{c}{\textit{Software Engineering Agent}} \\
\midrule
\multirow{1}{*}{SWE-agent} 
  & \cellcolor{gray!10} DeepSeek-V3 & 15.5 & 29.3 & 56.7 & 51.8 & 9.7 \\
\midrule
\multicolumn{7}{c}{\textit{Environment Configuration Agent}} \\
\midrule
\multirow{1}{*}{Installamatic}
  & \cellcolor{gray!10} DeepSeek-V3 & 6.7 & / & / & / & / \\
\multirow{1}{*}{Repo2Run} 
  & \cellcolor{gray!10} DeepSeek-V3 & 44.8 & / & / & / & / \\
\multirow{1}{*}{\ours} 
  & \cellcolor{gray!10} DeepSeek-V3 & \textbf{63.2} & \textbf{41.3} & \textbf{98.7} & \textbf{68.7} & \textbf{71.7} \\
\bottomrule
\end{tabular}
}
\end{table}

\begin{table}[!h]
\centering
\caption{Performance across different scenarios on Python repositories from \oursbench using DeepSeek-V3. \textbf{Bold} indicates the best performance.}
\label{tab:rat-python-difficulty}
\renewcommand{\arraystretch}{1.1} 
\setlength{\tabcolsep}{7pt} 
\begin{tabular}{lcccc}
\toprule
\textbf{Framework} & \textbf{S1} & \textbf{S2} & \textbf{S3} & \textbf{Avg.} \\
\midrule
Repo2Run & 39.8 & 25.4 & 3.0 & 22.7 \\
RAT & \textbf{50.5} & \textbf{69.5} & \textbf{92.0} & \textbf{70.7} \\
\bottomrule
\end{tabular}
\end{table}

\vpara{Ablation Studies.}To assess the contribution of each component in \ours, we perform ablation studies with the following variants: (1) \ours w/o init, without ImageRetriever for initialization; (2) \ours w/o tool, without specialized toolset; and (3) \ours-auto, with automated mode instead of standard mode. Details are included in Appendix~\ref{app:exp-setup}.

\begin{table}[!h]
\centering
\caption{Ablation study of \ours. Performance is reported with the DeepSeek-V3 model on Python repositories. \textbf{Bold} indicates best performance.} 
\label{tab:ablation}
\renewcommand{\arraystretch}{1.3} 
\setlength{\tabcolsep}{8pt} 
\resizebox{\columnwidth}{!}{
\begin{tabular}{lccc}
\toprule
\textbf{Variant} & \textbf{ESSR (\%)} & \textbf{Tokens (K)} & \textbf{Latency (min)} \\
\midrule
\ours w/o init & 40.5 & \textbf{180.8} & 18.3 \\
\ours w/o tool & 55.7 & 351.2 & 36.9\\
\ours-auto & 56.9 & 364.2 & \textbf{16.0} \\
\midrule 
\ours & \textbf{63.2} & 421.9 & 24.3 \\
\bottomrule
\end{tabular}
}
\end{table}

As shown in Table~\ref{tab:ablation}, removing the ImageRetriever (\ours w/o init) or the specialized toolset (\ours w/o tool) results in drop in ESSR, indicating that high-quality initial images and precise tool execution are key drivers of success. Additionally, while \ours consumes the most tokens, it maintains a competitive latency of 24.3 minutes, significantly faster than \ours w/o tool variants, demonstrating that \ours achieves an effective balance between high success rates and configuration efficiency. Regarding configuration modes, the automatic mode improves flexibility by eliminating fixed workflows, substantially reducing configuration time and token usage, with only a modest performance drop compared to the standard mode.

\vpara{Performance under Different Backbones.}
We evaluate the performance of \ours across a range of backbone models. Following the code capability rankings reported in LMArena~\cite{chiang2024chatbot}, we select representative LLMs spanning different capability tiers, including relatively weaker backbones Qwen3-Coder-30B~\cite{yang2025qwen3}, and stronger models such as DeepSeek-V3, GLM-5, and GPT-5.2, as summarized in Table~\ref{tab:backbone-comparison}.

\begin{table}[!h]
\centering
\caption{Effect of backbone models on Repo2Run and \ours. Performance is reported on Python repositories from \oursbench. \textbf{Bold} indicates the best performance within each framework.}
\label{tab:backbone-comparison}
\small
\renewcommand{\arraystretch}{1.15}
\setlength{\tabcolsep}{4pt}

\begin{tabular}{llcc}
\toprule
\multicolumn{2}{c}{\textbf{Model Configuration}} & \multicolumn{2}{c}{\textbf{Metrics}} \\
\cmidrule(lr){1-2} \cmidrule(lr){3-4}
Framework & LLM & ESSR (\%) & Tokens (K) \\
\midrule
\multirow{4}{*}{Repo2Run} 
& \cellcolor{gray!5} Qwen3-Coder-30B & 33.2 & \textasciitilde{} 350 \\
& \cellcolor{gray!10} DeepSeek-V3 & 44.8 & \textasciitilde{} 400 \\
& \cellcolor{gray!15} GLM-5 & \textbf{59.3} &  \textasciitilde{} 400 \\
& \cellcolor{gray!20} GPT-5.2 & 25.1 & \textasciitilde{} 350 \\
\midrule

\multirow{4}{*}{RAT}
& \cellcolor{gray!5} Qwen3-Coder-30B & 47.2 & 355.6 \\
& \cellcolor{gray!10} DeepSeek-V3 & 63.2 & 421.9 \\
& \cellcolor{gray!15} GLM-5 & 69.5 & 264.3 \\
& \cellcolor{gray!20} GPT-5.2 & \textbf{86.8} & 280.2 \\
\bottomrule
\end{tabular}
\end{table}

As shown in Table~\ref{tab:backbone-comparison}, \ours consistently outperforms Repo2Run across all backbone settings, even with weaker models (e.g., Qwen3-Coder-30B). Notably, \ours shows improved performance as stronger backbones are adopted, with GPT-5.2 achieving the best results. This trend is consistent across both Repo2Run and RAT, highlighting the robustness and strong generalization of our framework with respect to backbone selection.

\vpara{Comparison with Human Engineers.}
To evaluate the performance of automated tools, we engaged several senior engineers to manually configure environments for three representative Python repositories. The experiment followed a standardized protocol: engineers first selected an appropriate Python base image guided by the repository's README. They then cloned the target GitHub repositories and checked out specific commits. The primary goal was to successfully execute \texttt{pytest --collect-only -q} and run internal tests via \texttt{pytest -q}. We recorded the total latency and ESSR to quantify manual overhead. Additionally, each task was assessed on a 5 point Likert scale along two dimensions of cognitive workload: Difficulty and Effort. Detailed experiment settings and more extensive experiments on repositories of other programming languages are provided in Appendix~\ref{app:exp-setup} and Appendix~\ref{app:exp}.

\begin{table}[!h]
\centering
\caption{Comparison of Python environment configuration between senior human engineers and \ours. `/' denotes not applicable.}
\label{tab:human-baseline-results}
\renewcommand{\arraystretch}{1.3} 
\setlength{\tabcolsep}{6pt}
\resizebox{\columnwidth}{!}{
\begin{tabular}{lcccc}
\toprule
\textbf{Group} & \multicolumn{2}{c}{\textbf{Efficiency}} & \multicolumn{2}{c}{\textbf{Cognitive Load}} \\
\cmidrule(lr){1-1} \cmidrule(lr){2-3} \cmidrule(lr){4-5}
Role & ESSR (\%) & Latency (min) & Difficulty & Effort \\
\midrule
Engineers & 89.41 & \textbf{13.73} & 2.9 / 5.0 & 3.3 / 5 \\
\ours  & \textbf{91.52} & 31.26 & / & / \\
\bottomrule
\end{tabular}
}
\end{table}

\begin{figure*}[!t]
    \centering
    \includegraphics[width=\textwidth]{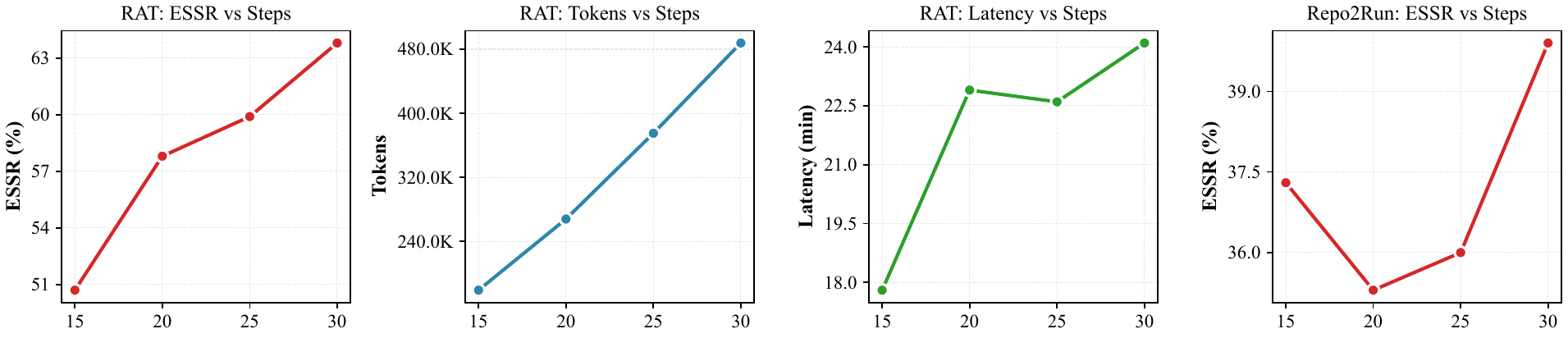}
    \caption{Performance of \ours and Repo2Run under varying execution step budgets. The first three panels correspond to \ours, and the last panel shows Repo2Run. As the step budget increases, \ours consistently improves ESSR, at the cost of higher token usage and latency.}
    \label{fig:scaling_effects}
\end{figure*}

\begin{figure*}[!t]
    \centering
    \includegraphics[width=\textwidth]{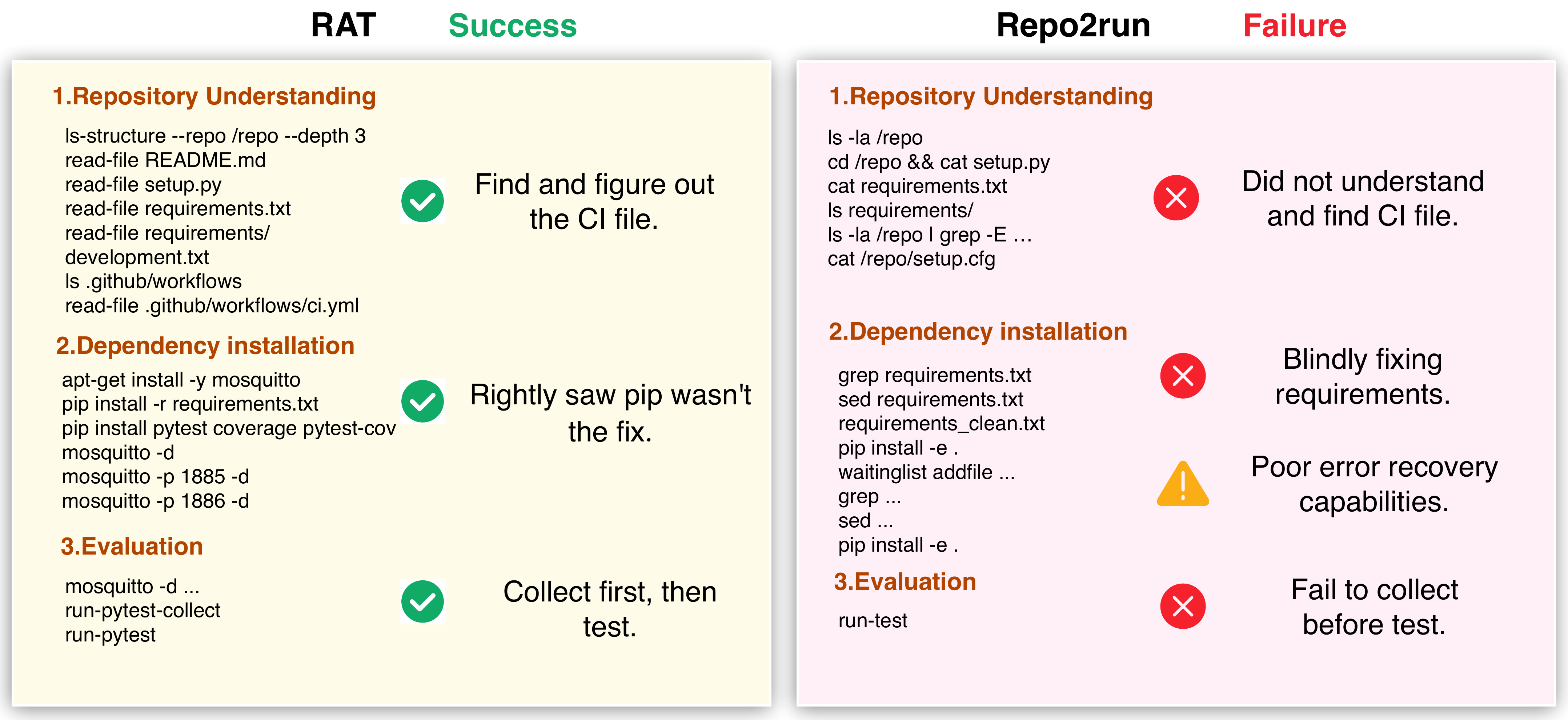}
    \caption{Trajectory comparison between \ours and Repo2Run on repository stlehmann/Flask-MQTT.}
    \label{fig:trajectory}
\end{figure*}

As shown in Table~\ref{tab:human-baseline-results}, although \ours is slower than human engineers in configuration time by a factor of 2.25, it achieves a 2.36\% relatively higher ESSR. This demonstrates the effectiveness of our agent, which can even surpass engineers in environment configuration. Moreover, \ours requires no human intervention and supports parallelized environment setup across large-scale repositories, where manual effort would be costly. Thus, additional time overhead is acceptable given its efficiency and superior performance.

\begin{figure}[!ht]
    \centering
    \includegraphics[width=0.50\textwidth]{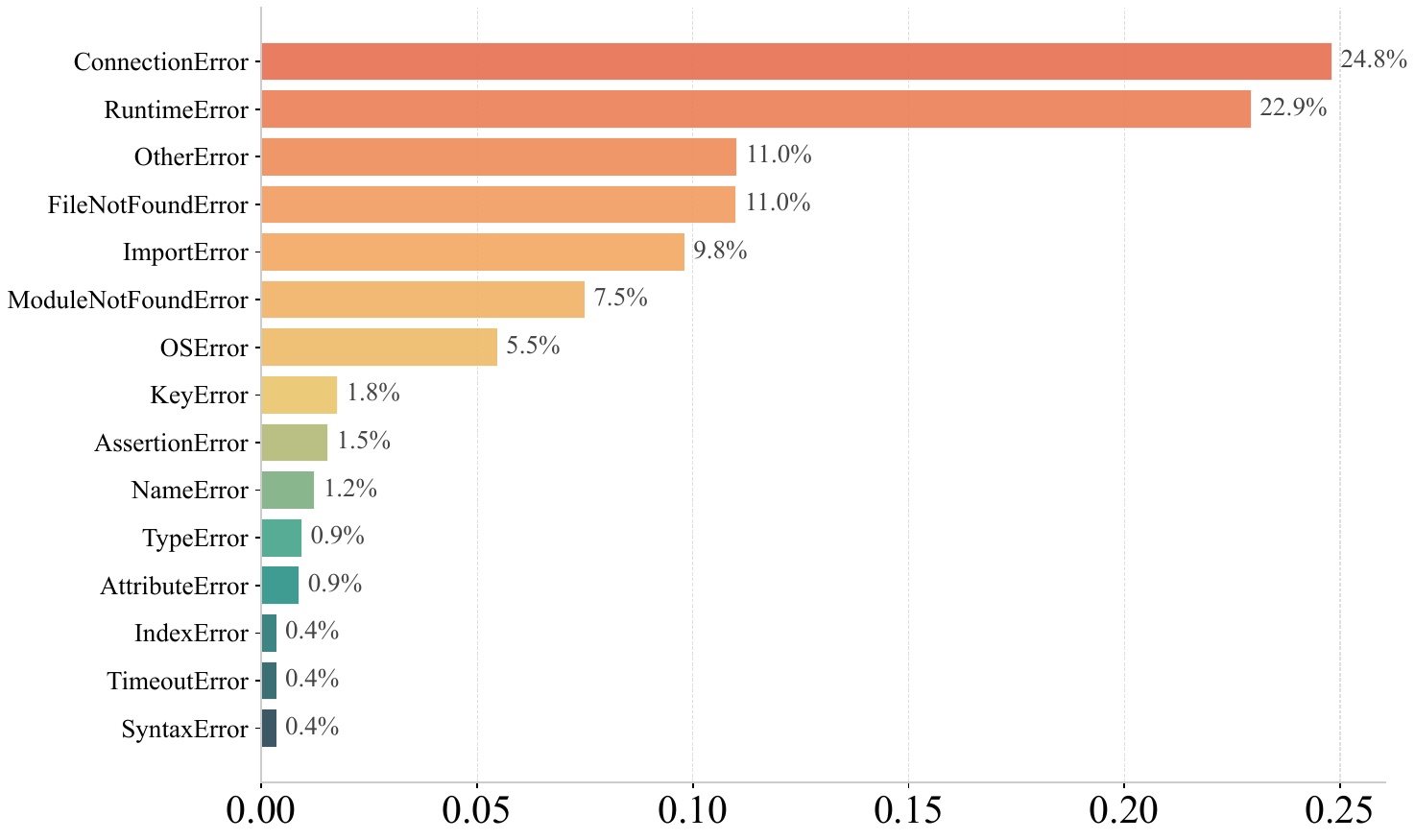}
    \caption{Breakdown of pytest error types for Python repositories where \ours fails to solve.}
    \label{fig:error_breakdown}
\end{figure}

\vpara{Failure Analysis.} Figure~\ref{fig:error_breakdown} summarizes the dominant error categories among Python repositories where \ours fails to complete verification. We identify two major failure modes: (1) \emph{ConnectionError}, which mainly occurs in API-reliant repositories when tests fail to reach external services (e.g., APIs, databases, or brokers). These errors stem from restricted network access, missing credentials, or unprovisioned local services; and (2) \emph{RuntimeError}, which occurs after installation succeeds and is typically caused by missing system libraries, incompatible binaries, or hardware/driver assumptions. These failures are difficult to repair automatically because they often require OS-level dependencies, platform-specific configurations, or unavailable external resources. More broadly, LLM-driven environment configuration faces a practical ceiling when repositories are inherently problematic, setup information is incomplete or ambiguous, dependencies are no longer reproducible, or execution relies on private APIs, credentials, or proprietary data, introducing uncertainty that cannot be resolved solely through reasoning.

\vpara{Scaling Effects in Configuration.}
We further evaluate ESSR under varying execution steps. As shown in Figure~\ref{fig:scaling_effects}, \ours consistently improves success rate with increasing steps, demonstrating a clear scaling-law like behavior similar to LLM reasoning~\cite{wei2022chain,wu2024inference,yuan2026tree}. In contrast, Repo2Run shows no stable improvement, indicating that our framework enables more predictable compute–performance scaling in environment configuration.

Furthermore, the growth of average latency slows down over time, implying a better balance between exploration and exploitation in later stages. This indicates that additional steps mainly refine solutions rather than proportionally increasing computational overhead.

\vpara{Case Study on Trajectories.} We take \texttt{stlehmann/Flask-MQTT} as an example repository to illustrate the different trajectory between \ours and Repo2Run as shown in Figure~\ref{fig:trajectory}. Compared to Repo2Run, our agent explicitly aligns its configuration strategy with the repository’s CI workflow and runtime requirements. By inspecting CI scripts, \ours correctly identifies system-level service dependencies (e.g., Mosquitto brokers) that are invisible to Python-centric dependency analysis. This enables our agent to provision the execution environment holistically before test execution, whereas Repo2Run repeatedly attempts to resolve failures through requirement-level manipulations, leading to non-convergent behavior.
\section{Related Work}
\subsection{Code agents.} Autonomous software engineering has evolved from static generation to reasoning-driven agents capable of repository-level problem solving. Early works like MetaGPT~\cite{hong2023metagpt} introduced SOP-based multi-agent collaboration, while CodeChain~\cite{le2023codechain} leveraged modular self-revisions. Recent systems like SWE-agent~\cite{yang2024swe} optimize the Agent-Computer Interface (ACI) for benchmarks like SWE-bench~\cite{jimenez2023swe}. Beyond prompt or workflow engineering, training-based frameworks such as Agent-RLVR~\cite{da2025agent} employ environment-based rewards and pedagogical guidance to refine software engineering trajectories. Complementary work explores improved exploration and policy learning for agents~\cite{zhang2026rapo}, alongside sandbox‑based simulation~\cite{huang2026policysim,gao-etal-2026-towards}. However, despite their proficiency in patch generation, these agents typically assume pre-configured environments, leaving autonomous environment configuration largely unaddressed.

\subsection{Environment configuration.} The research focus for repository-level tasks has shifted from isolated code generation to environment configuration. Alongside the rapid increase of generic and domain-specific benchmarks~\cite{hua2026ptcg,tang2026chartwalker} for agents, early benchmarks like SWE-Bench~\cite{jimenez2023swe} identified real-world resolution difficulties, they were hindered by manual setup requirements. Recent benchmarks such as EnvBench \cite{eliseeva2025envbench}, GitTaskBench \cite{ni2025gittaskbench}, and systematic analyses of Python ecosystem \cite{milliken2025beyond} have addressed it by providing workflows and ground-truth installation processes, establishing environment setup as a cornerstone of autonomous software engineering. However, existing benchmarks only offer preliminary explorations of configuration and fail to reflect the distribution of real-world repositories.

To tackle environment configuration, recent studies employ agentic strategies with iterative feedback. For instance, Repo2Run~\cite{hu2025repo2run} and ExecutionAgent~\cite{bouzenia2025you} use LLM reasoning to synthesize Dockerfiles and refine scripts across languages based on execution outcomes. Furthermore, SETUPAGENT~\cite{ni2025gittaskbench} automates benchmark construction, enabling large-scale datasets. These advancements mark a shift from rule-based installation to dynamic, reasoning-driven agents for complex software dependencies. However, current methods still rely on configuration artifacts or are constrained by language-specific limitations, and thus cannot handle generalized environment configurations.

\section{Conclusion}
Environment configuration is a bottleneck for autonomous code agents. To address this problem, we introduce \ours (RunAnyThing), a novel modular and extensible framework for fully automated environment configuration across programming languages. By integrating semantic initialization with language-aware abstraction, configuration mode and configuration toolset, \ours effectively mitigates information sparsity and resolves complex dependencies. To enable comprehensive evaluation on real-world repositories, we construct \oursbench, a multilingual benchmark comprising over 2,500 real-world repositories. Experimental results demonstrate that \ours achieves state-of-the-art performance among existing baselines and approaches the setup success rates of senior human engineers. Future work will focus on scaling \ours to a wider range of repositories and more complex deployment scenarios.

\section*{Limitations}
Although \ours can effectively construct executable environments for many repositories, it still has limitations. (1) Our benchmark and pipeline assume a single-container setting, which simplifies evaluation but does not cover common multi-service deployments (e.g., via \texttt{docker-compose}) involving cross-container networking, service readiness, shared volumes, and versioned sidecars; Docker-in-Docker further complicates this issue. (2) In addition, hardware-dependent environments (e.g., GPU workloads requiring strict alignment of drivers, runtimes, and CUDA/cuDNN) remain out of scope due to their complex and hard-to-reproduce failures. (3) Finally, some configurations require external human-provided inputs such as API keys or credentials, which are not currently handled. We leave multi-service, hardware-aware, and human-in-the-loop configuration to future work, with detailed error analysis in Appendix~\ref{app:exp}.

\section*{Ethics Considerations}
As an autonomous agent capable of executing arbitrary repository code, RAT raises ethical and security concerns. Automated execution of unverified code may introduce risks such as malicious scripts, remote code execution, or supply chain attacks. To mitigate these risks, RAT executes repositories in isolated Docker-based sandboxes and validates generated environments before execution. While container isolation provides a security boundary, production deployments should further adopt privilege restrictions (e.g., rootless containers and syscall filtering), resource limits (e.g., CPU, memory, and GPU quotas), and network isolation with allowlists. Dependency management also requires careful handling, including registry pinning, network whitelisting, or internal mirrors, to prevent attacks such as typosquatting and dependency confusion. In this work, AI-assisted writing is used solely for refining descriptions and does not affect system design or execution logic.

\section*{Acknowledgements}
This work is supported by NSFC (No.\ 62322606, No.\ 62441605, No.\ 62606466), and is sponsored by CAAI-Ant Group Research Fund.


\bibliography{custom}

\newpage
\clearpage
\appendix
\section{Notations}\label{app:notation}
The main notations are summarized in Table~\ref{tab:notations}.
\begin{table*}[t]
\centering
\fontsize{9.8pt}{11.8pt}\selectfont
\setlength{\tabcolsep}{3pt}
\renewcommand{\arraystretch}{1.15}
\caption{\label{tab:notations} Description of major notations.}
 \begin{tabular}{@{}p{0.23\textwidth}p{0.70\textwidth}@{}}
		\toprule 
		\textbf{Notation} & \textbf{Description}   \\
		\midrule
        $\mathcal{R}$ & A repository comprising source code and metadata. \\
        $\mathcal{E}(\mathcal{R})$ & The set of feasible containerized environments compatible with repository $R$. \\
        $e \in \mathcal{E}(\mathcal{R})$ & A specific environment instance constructed for the task. \\
        $\pi(\mathcal{R})$ & A repository-dependent verification sequence (e.g., tests or build commands). \\
        $\tau(\mathcal{R})$ & An interaction trace defined as a sequence of action-observation pairs $((a_1, o_1), ..., (a_T, o_T))$. \\
        $a_t, o_t$ & The tool invocation (action) and resulting system feedback (observation) at time step $t$. \\
        $N$ & The total number of unit tests available in a given repository. \\
        $N_{\text{pass}}$ & The number of unit tests that pass successfully within the configured environment. \\
        $N_{\text{verified}}$ & The number of verified ground-truth tests. \\
		\bottomrule 
	\end{tabular}
\end{table*}

\section{Framework} \label{app:framework}
The pseudocode of the algorithm underlying \ours is presented in Algorithm~\ref{alg:env-setup}.
\begin{figure*}[t]
\centering
\begin{minipage}{0.96\textwidth}
\begin{algorithm}[H]
\caption{Automated Environment Construction in \ours (Standard Plan Mode)}
\label{alg:env-setup}
\begin{algorithmic}[1]
\REQUIRE Repository $\mathcal{R}$, maximum turns $T$,  LLM backbone
\ENSURE Environment $e \in \mathcal{E}(\mathcal{R})$, Interaction trace $\tau(\mathcal{R})$, Dockerfile $F$
\STATE \textbf{SetupAgent:}
\STATE \quad Extract configuration artifacts $\mathcal{C} \subset \mathcal{R}$ and identify primary language $\mathcal{L}$
\STATE \quad \textbf{ImageRetriever:}
\STATE \quad \quad Perform semantic analysis on $\mathcal{C}$ and recommend base image $I$  from the default image set
\STATE \quad \quad \textbf{if} LLM determines to search Docker Hub \textbf{then}
\STATE \quad \quad \quad Query specialized image set $\mathcal{I}_{hub}$ via Docker Hub
\STATE \quad \quad \quad Execute LLM-based scoring and select best base image $I \leftarrow score\_select(I,\mathcal{I}_{hub})$
\STATE \quad \quad \textbf{end if}
\STATE \quad Construct initial Dockerfile $F_0$ for $e$ using a template based on $\mathcal{L}$ and $I$
\STATE \textbf{Image Build and Validation:} Validate $e$ in a temporary context with fallback mechanisms
\STATE \textbf{Environment Instantiation:} Create container from $e$, inject specialized toolset $\mathcal{S}_{tool}(\mathcal{L})$

\FOR{$t = 1$ to $T$}
    \STATE \textbf{ReAct Loop:}
    \STATE \quad \textbf{Thought:} $\text{LLM}$ reasons over repository $\mathcal{R}$ and trajectory $\tau(\mathcal{R})$
    \STATE \quad \textbf{Action:} Select and invoke $a_t \in \mathcal{S}_{tool} \cup \text{BASH}$
    \STATE \quad \textbf{Observation:} Capture system feedback $o_t$ (stdout, stderr, tool output)
    \STATE Update trajectory $\tau(\mathcal{R}) \leftarrow \tau(\mathcal{R}) \cup \{(a_t, o_t)\}$
    
    \IF{configuration $\text{success}$ \OR $\text{critical failure}$}
        \STATE \textbf{break}
    \ENDIF
\ENDFOR
\STATE LLM infers Dockerfile $F$ from $F_0 \cup \tau(\mathcal{R})$
\STATE \textbf{return} $e, \tau(\mathcal{R}),F$
\end{algorithmic}
\end{algorithm}
\end{minipage}
\end{figure*}

\section{\oursbench Details}\label{app:bench_stats}
In this section, we report summary statistics of \oursbench to characterize the diversity introduced by the benchmark construction procedure. All statistics and visualizations are computed on a balanced core split of 2,500 repositories (500 per language) spanning Python, Java, JavaScript/TypeScript, Go and Rust.

\vpara{Language Coverage and Verification.}
\oursbench spans five programming languages, each with distinct environment configuration characteristics.
For Java, Rust, Go, and JS/TS, we rely on standard build manifests and verify setups via deterministic build or test commands. In contrast, Python environments are validated through unit tests, supplemented by Dockerfiles or CI scripts (if available), to ensure reliable verification. Table~\ref{tab:language_characteristics} summarizes these language-specific inclusion signals and their dominant failure modes.

\begin{table*}[t]
\centering
\caption{Detailed language-specific characteristics, inclusion signals, and dominant failure modes in \oursbench.}
\label{tab:language_characteristics}
\fontsize{9.8pt}{11.8pt}\selectfont
\setlength{\tabcolsep}{3pt}
\renewcommand{\arraystretch}{1.2}
\begin{tabular}{@{}lccccc@{}}
\toprule
\textbf{Feature} & \textbf{Python} & \textbf{Java} & \textbf{Rust} & \textbf{JS/TS} & \textbf{Go} \\
\midrule
\textbf{Required manifest} & -- & \makecell[l]{\texttt{pom.xml} / \\ \texttt{build.gradle}} & \texttt{Cargo.toml} & \texttt{package.json} & \texttt{go.mod} \\
\addlinespace[4pt]
\textbf{Verification} & \texttt{pytest} & \makecell[l]{\texttt{mvn install} / \\ \texttt{gradle build}} & \makecell[l]{\texttt{cargo build} / \\ \texttt{cargo test}} & \makecell[l]{\texttt{npm install} / \\ \texttt{npm test}} & \makecell[l]{\texttt{go build} / \\ \texttt{go test}} \\
\addlinespace[4pt]
\textbf{Unique feature} & \makecell[l]{Dynamic runtime \\ dependencies} & Build lifecycle & \makecell[l]{Compiler safety \\ guarantees} & Transpilation & \makecell[l]{Module-aware \\ compilation} \\
\addlinespace[4pt]
\textbf{Primary risk} & \makecell[l]{Missing system \\ libraries} & \makecell[l]{Dependency \\ version conflicts} & \makecell[l]{Linker or target \\ mismatch} & \makecell[l]{Node.js or native \\ module issues} & \makecell[l]{CGO or module \\ resolution issues} \\
\bottomrule
\end{tabular}
\end{table*}

\begin{figure}[t]
    \centering
    \includegraphics[width=\linewidth]{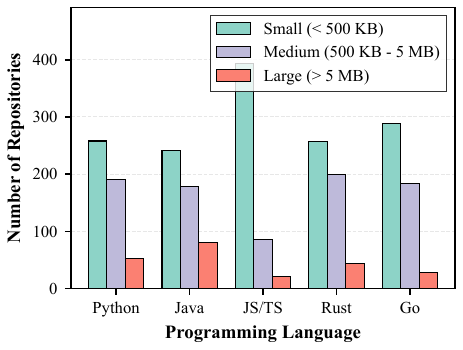}
    \caption{Repository size distribution across languages in \oursbench.}
    \label{fig:ratbench_size_distribution}
\end{figure}

\vpara{Repository Size Distribution.} Repository size is measured by code bytes (excluding non-code assets such as images and documentation) and discretized into three tiers: Small ($<500$ KB), Medium (500 KB--5 MB), and Large ($>5$ MB). Figure~\ref{fig:ratbench_size_distribution} and Table~\ref{tab:ratbench_size_counts} summarize size-tier compositions by language. The distributions differ substantially across ecosystems: JS/TS is dominated by small repositories, while Java contains a noticeably larger fraction of large repositories. This heterogeneity is important for environment configuration because large projects tend to introduce deeper build-tool stacks (e.g., multi-module builds and native toolchains), whereas smaller packages often stress dependency resolution and versioning behavior in language-specific package managers.

\begin{table}[t]
    \centering
        \caption{Repository size distribution by language on the 2,500 repository core split.}
        \label{tab:ratbench_size_counts}
        \setlength{\tabcolsep}{3pt}
        \renewcommand{\arraystretch}{1.2}
        \begin{tabular}{lccc}
            \toprule
            \textbf{Language} & \textbf{Small} & \textbf{Medium} & \textbf{Large} \\
            \midrule
            Python  & 258 (51.6\%) & 190 (38.0\%) & 52 (10.4\%) \\
            Java    & 242 (48.4\%) & 178 (35.6\%) & 80 (16.0\%) \\
            JS/TS   & 393 (78.6\%) & 86 (17.2\%)  & 21 (4.2\%)  \\
            Rust    & 257 (51.4\%) & 199 (39.8\%) & 44 (8.8\%)  \\
            Go      & 289 (57.8\%) & 183 (36.6\%) & 28 (5.6\%)  \\
            \bottomrule
        \end{tabular}
\end{table}

\begin{figure}[t]
    \centering
    \includegraphics[width=\linewidth]{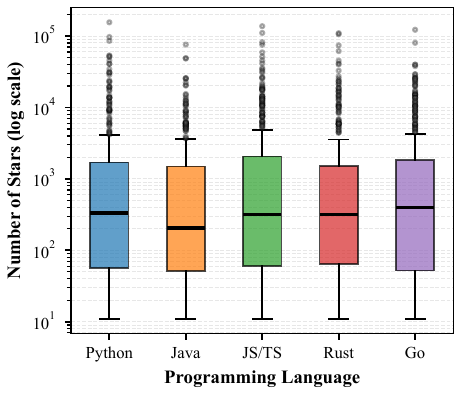}
    \caption{Repository popularity (GitHub stars) distribution by language in \oursbench.}
    \label{fig:ratbench_popularity_distribution}
\end{figure}

\vpara{Repository Popularity Distribution.} Repository popularity is measured by GitHub stars. Figure~\ref{fig:ratbench_popularity_distribution} and Table~\ref{tab:ratbench_star_stats} summarize the star distributions by language. Popularity is long-tailed (range: 11 to 155,569 stars), with a median 294 and a mean 2,868, consistent with typical open-source ecosystems. While stars are not a direct proxy for configuration complexity, including both long-tail and top-tier projects helps prevent benchmark bias toward either toy repositories (often under-documented) or highly engineered projects (often with mature CI/CD and containerization).

\begin{table}[t]
    \centering
    \caption{GitHub stars summary statistics by language on the 2,500 repository core split.}
    \label{tab:ratbench_star_stats}
    \setlength{\tabcolsep}{3pt}
    \renewcommand{\arraystretch}{1.2}
    \begin{tabular}{lcccc}
        \toprule
        \textbf{Language} & \textbf{Min} & \textbf{Max} & \textbf{Median} & \textbf{Mean} \\
        \midrule
        Python  & 11 & 155{,}569 & 333 & 3{,}163 \\
        Java    & 11 & 75{,}937  & 205 & 1{,}880 \\
        JS/TS   & 11 & 137{,}288 & 316 & 3{,}564 \\
        Rust    & 11 & 109{,}632 & 321 & 2{,}864 \\
        Go      & 11 & 122{,}421 & 398 & 3{,}002 \\
        \bottomrule
    \end{tabular}
\end{table}

\section{Additional Experimental Setup}\label{app:exp-setup}

\vpara{Implementation Details.} We detail the implementation of evaluated models below. For \ours, we set the LLM temperature to 0.0 for reproducibility and limit each session to 30 turns for efficiency. The system employs a multi-layered timeout strategy (600s for commands; 7200s global) and allocates 150K token limit to support resource-intensive builds. 

All experiments are conducted on a high-performance Linux server equipped with dual-socket AMD EPYC 9654 processors, totaling 224 CPU cores  with 2 hardware threads per core. The system provides shared-memory capacity with two NUMA nodes, supporting efficient parallel execution for large-scale environment configuration workloads.

\vpara{Description of Baselines.} To ensure fair and consistent evaluation, all baseline methods implement a unified interface that adapts to our evaluation framework. Specifically, each baseline takes a repository name as input and outputs a properly configured container environment. This standardized interface enables us to systematically run the corresponding test runners within each container to evaluate setup success across all methods. Below, we provide descriptions of the implementation of different baselines used in our experiments.
\begin{itemize}[leftmargin=10pt,nosep]

\item \textbf{pipreqs.} This baseline employs a static analysis approach and generating a container using predefined Dockerfile templates. It represents a deterministic, non-learning baseline that relies solely on manifest-based dependency resolution without dynamic adaptation. The Dockerfile template used for this baseline is provided in Section~\ref{app:prompts}.

\item \textbf{Zero-Shot.} Unlike template-based methods, this baseline directly prompts an LLM to generate a complete Dockerfile from scratch based on repository analysis. It evaluates the model's ability to perform environment configuration in a single-shot generation without iterative refinement or tool-assisted interaction. The prompt template is provided in Section~\ref{app:prompts}.

\item \textbf{SWE-agent.} We adapt SWE-agent to the environment configuration task by customizing its system prompts, agent workflow, and tool configurations to align with our evaluation framework. The complete configuration is provided in Section~\ref{app:prompts} for reproducibility.

\item \textbf{Installamatic.} To ensure a consistent evaluation environment, we adapt the original Installamatic repository to run fully locally on Linux, removing the need for a virtual machine and enabling direct Docker-based evaluation. For LLM consistency, we re-implement the LLM inference interface to support the LLM API. Key changes include local Docker execution, API modification, and minor initialization updates.

\item \textbf{Repo2Run.} Repo2Run follows an agent-based framework architecturally similar to our evaluation setup. We apply minimal workarounds to adapt its interface to our benchmark, enabling direct integration with our standardized evaluation pipeline without major structural modifications.
\end{itemize}

\vpara{Settings for ablation studies.} Here, we elaborate in detail on how each ablation study is conducted:
\begin{itemize}[leftmargin=10pt,nosep]
\item \textbf{\ours w/o init.} The ImageRetriever module is deactivated. Instead of performing semantic analysis to infer and retrieve the most suitable Docker image from Docker Hub, this variant initializes the sandbox with a fixed, default base image corresponding to the identified primary programming language (e.g., \texttt{python:3.10-slim} for Python). This variant verifies the necessity of retrieving project-specific runtime environments for robust initialization.
\item \textbf{\ours w/o tool.} The specialized agent toolset is disabled. The agent is restricted to interacting with the environment solely through basic shell commands (e.g., \texttt{grep}, \texttt{sed}, \texttt{cat}, and \texttt{echo}) and essential tools (e.g., \texttt{STOP}), lacking access to the high-level capabilities such as web search or issue retrieval. This variant verifies the contribution of the specialized tool abstractions to the configuration precision and efficiency.
\item \textbf{\ours-auto.} This variant corresponds to the automated mode. Agent would interacts autonomously to discover repository-specific workflows. This design enables structured decomposition of complex configuration tasks and supports dynamic adaptation during the configuration process.
\end{itemize}

\vpara{Settings for Table~\ref{tab:backbone-comparison}.}
To ensure a controlled backbone comparison, \ours and Repo2Run use the same hyper-parameters: each session is capped at 30 turns. We also set the context budget to 150K tokens. We evaluate \emph{weaker backbones} on 150 Python repositories, while \emph{stronger backbones} are evaluated on a smaller 30 Python repositories subset due to cost considerations. Notably, Repo2Run does not expose a token-accounting interface, so we cannot report its exact token consumption.

\vpara{Settings for Comparison with Human Engineers.} We recruited several senior engineers to participate in the experiments, with informed consent obtained from all participants. To quantify the manual overhead beyond time and success rates, we adopted a subjective assessment framework. Upon completion of each environment configuration task, senior engineers were required to rate their experience based on two cognitive dimensions:

\begin{itemize}[leftmargin=10pt,nosep]
    \item \emph{Difficulty}: Measures the technical complexity and the presence of obstacles (e.g., dependency conflicts, vague documentation) encountered during the setup.
    \item \emph{Effort}: Measures the mental and physical energy required to complete the task, reflecting the intensity of the engineer's involvement.
\end{itemize}

The evaluation utilized a 5-point Likert scale as Table~\ref{tab:likert_definition}, ranging from 1 (Very Low/Easy) to 5 (Very High/Difficult). This dual-metric approach allows us to distinguish between tasks that are technically complex but routine (High Difficulty, Moderate Effort) and those that are tedious and draining (Moderate Difficulty, High Effort).

\begin{table*}[t]
\centering
\caption{The 5-point Likert scale for Difficulty and Effort assessment.}
\label{tab:likert_definition}
\setlength{\tabcolsep}{4pt}
{\renewcommand{\arraystretch}{1.2}
\begin{tabular}{c l p{0.36\textwidth} p{0.34\textwidth}}
\toprule
\textbf{Score} & \textbf{Level} & \textbf{Difficulty (Technical)} & \textbf{Effort (Cognitive)} \\ \midrule
1 & Very Low & Straightforward; follows README perfectly. & Minimal mental energy required. \\
2 & Low & Minor tweaks or version adjustments needed. & Slight focus; routine operations. \\
3 & Moderate & Requires external search or troubleshooting. & Sustained attention; moderate fatigue. \\
4 & High & Major conflicts; requires deep debugging. & High mental strain; multiple attempts. \\
5 & Very High & Severe blockers; requires manual code fix. & Exhausting; requires extreme persistence. \\ \bottomrule
\end{tabular}
}
\end{table*}

In addition, the repositories we tested include projects from different domains, such as \texttt{opengeos/segment-geospatial}, \texttt{dsphper/lanhu-mcp}, and \texttt{python-escpos/python-escpos}. We then report the averages over these repositories.

\section{Specifications of Tools}\label{app:tool_specifications}
\ours supports a suite of specialized tools within the \ours framework. Table~\ref{tab:tool_specifications} and following outlines the comprehensive inventory of these tools and their functional specifications.

\begin{table*}[t]
\centering
\renewcommand{\arraystretch}{1.2}
\caption{Functional specifications of the tools provided in \ours.}
\label{tab:tool_specifications}
\begin{tabular}{p{0.23\textwidth} >{\raggedright\arraybackslash}p{0.70\textwidth}}
    \toprule
    \textbf{Tool Name} & \textbf{Functional Description} \\ \midrule
    \multicolumn{2}{l}{\textit{\textbf{Repository Analysis}}} \\ \nopagebreak
    \texttt{construct\_test} & Scans the repository to identify entry points, extracts run commands from READMEs, and locates test modules. The output includes entry point, run commands, and test information. \\
    \texttt{ls\_structure} & Displays the repository directory tree with a configurable depth. Highlights important files such as \texttt{README}, \texttt{setup.py}, and \texttt{Dockerfile}. \\
    \texttt{view\_outline} & Extracts code outlines including classes and function signatures for a given file or directory. Supports recursive scanning and optional line numbers. \\
    \texttt{read\_file} & Performs file reading with optional LLM-guided analysis. Provides a deeper understanding of key components and dependencies compared to standard \texttt{cat}. \\
    \addlinespace[0.6em]
    
    \multicolumn{2}{l}{\textit{\textbf{Knowledge Retrieval}}} \\ \nopagebreak
    \texttt{search\_repo} & Conducts global code snippet searches. Supports multiple modes (detailed/simple/LLM) and provides paths, line numbers, and brief contextual notes. \\
    \texttt{search\_web} & Queries external sources like StackOverflow, GitHub, and official documentation for error resolution or general ``how-to'' guidance. \\
    \texttt{retrieve\_image} & Infers project requirements (e.g., \texttt{package.json}) to search Docker Hub and recommend relevant images and tags with pull commands. \\
    \texttt{retrieve\_issue} & Searches an issue database for solutions to specific error messages (e.g., \texttt{ModuleNotFoundError}) and suggests potential fixes. \\
    \addlinespace[0.6em]

    \multicolumn{2}{l}{\textit{\textbf{Environment Setup}}} \\ \nopagebreak
    \texttt{edit\_file} & Modifies file content using various modes (replace/insert/search/LLM). Includes regex support and automatic \texttt{.bak} backup creation. \\
    \texttt{detect\_environment} & Reports basic system information, including GPU availability, system OS, network status, and available mirrors or tools. \\
    \texttt{cicd\_config} & Analyzes GitHub CI/CD workflows (\texttt{.yml}) to generate setup scripts and command lists for environment replication. \\
    \addlinespace[0.6em]
    \texttt{change\_python\_version} & Switches the container's Python version. \textbf{Note:} This action resets the environment and discards all previously installed packages. \\
    \texttt{change\_java\_version} & Switches the container's Java version (e.g., 11, 17, 21). \textbf{Note:} Similar to Python versioning, this resets the current environment state. \\
    \texttt{stop} & Terminates the environment setup flow and ensures the current state is saved and logged. \\

    \multicolumn{2}{l}{\textit{\textbf{Validation}}} \\ \nopagebreak
    \texttt{run\_test} & Executes tests based on results from \texttt{construct\_test}. Supports different execution types, including test, run, and collect. \\
    \texttt{run\_pytest} & Automatically runs all \texttt{pytest} tests in the repository and categorizes errors (e.g., \texttt{ImportError}). Saves results to a structured JSON log. \\
    \texttt{run\_pytest\_collect} & Collects available \texttt{pytest} tests without executing them to detect import-time errors and count test cases. \\
    \addlinespace[0.6em]

    \bottomrule
\end{tabular}
\end{table*}

\begin{itemize}[leftmargin=15pt, nosep] 
\item \textbf{Read File:} Beyond a standard \texttt{cat} command, this tool enables LLM-powered semantic understanding of file's purpose, dependencies, and key logic patterns,  giving compact context for large files.

\item \textbf{Edit File:} Implements a GitHub-style diff mechanism to ensure precise modifications and supports line-range replacement, regex-based search-and-replace, as well as LLM-guided fuzzy matching. An automatic backup system prevents irreversible errors during iterative edits.

\item \textbf{View Outline:} Extracts function signatures, class definitions, and type annotations while filtering out implementation noise. It supports various programming languages (Python, JS/TS, Rust, Java) via AST-based parsing with regex fallback.

\item \textbf{Ls Structure:} Generates a filtered directory tree by pruning irrelevant artifacts, enabling the agent to focus on critical configuration entry points.

\item \textbf{Issue Retrieval:} When failures are project-specific, \ours queries an internal repository issue pool to reuse prior fixes. We first form a retrieval query from the observed error and an LLM-produced error synopsis, then rank candidates with a hybrid scorer that combines keywords and error types with an LLM reranker that judges semantic relevance and fix usefulness.

\item \textbf{Change Version:} This tool allows the agent to switch language versions within a container dynamically. By leveraging Docker commit to capture snapshots and enabling automated environment rollback, it provides a core utility for resolving version conflicts and syntax incompatibilities.

\item \textbf{Error Recovery:} Environment configuration is inherently error-prone, so an effective error recovery mechanism is vital for successful configuration. When encountering failures (e.g., \texttt{ModuleNotFoundError}), \ours leverages a multi-channel recovery solution: (1) utilizing the LLM's intrinsic debugging capabilities~\cite{majdoub2024debugging} to solve, (2) performing semantic search across the repository's historical issues to identify project-specific solutions, and (3) querying external knowledge information, such as Stack Overflow, to obtain community-documented fixes.

\item \textbf{Detect Environment:} An automated environment inspection tool designed to scan and report the configuration, capabilities, and available resources of the container.

\item \textbf{CI/CD Config:} A CI/CD configuration parser capable of automatically extracting environment configuration steps from GitHub Actions and converting them into executable commands for local containers.

\item \textbf{Test Synthesis:} For repositories lacking unit tests, \ours automatically generates lightweight smoke tests~\cite{chauhan2014smoke} to verify basic executability. It also supports entry-point scripts, validating deployment success through strategic timeouts.

\item \textbf{Test Runners:} This tool provides language-specific test runners (e.g., \texttt{run\_pytest} for Python) that autonomously discover and execute project tests. The runners apply specialized parsing strategies (e.g., JUnit XML parsing and regex-based analysis) to extract results and categorize failure modes systematically.
\end{itemize}

\section{Additional Experimental Results}\label{app:exp}

\begin{figure*}[!ht]
    \centering
    \includegraphics[width=\textwidth]{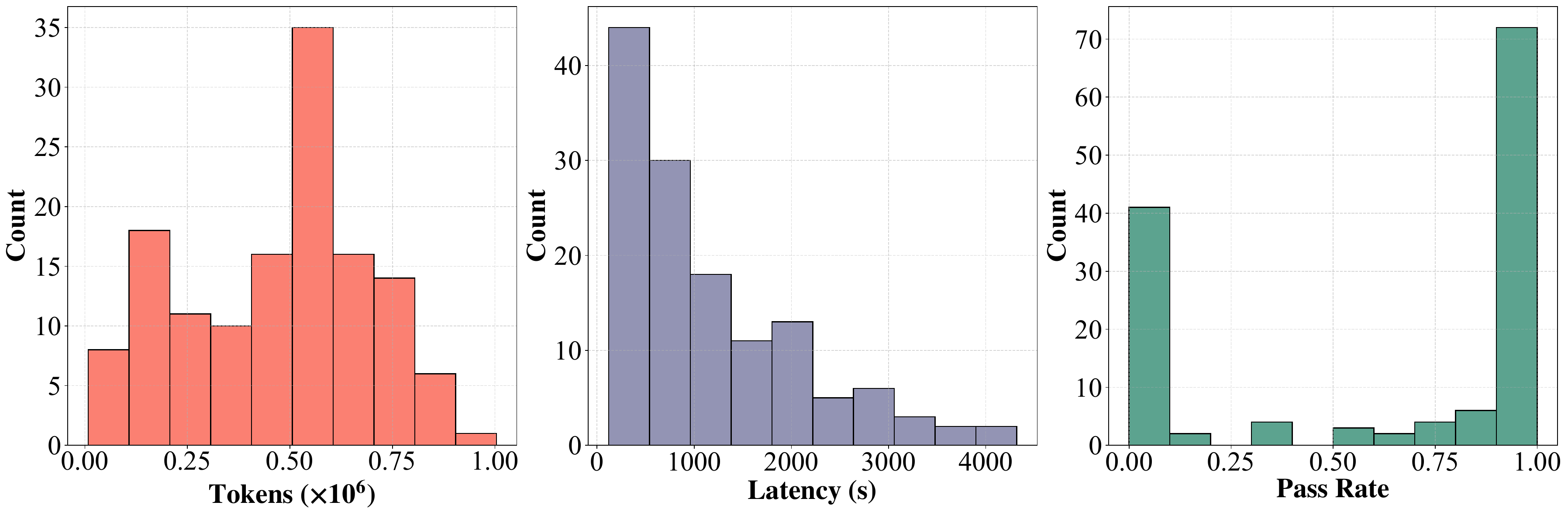}
    \caption{Distributions of tokens, latency, and pass rates across repositories.}
    \label{fig:distribution}
\end{figure*}

\begin{figure}[H]
    \centering
    \includegraphics[width=\linewidth]{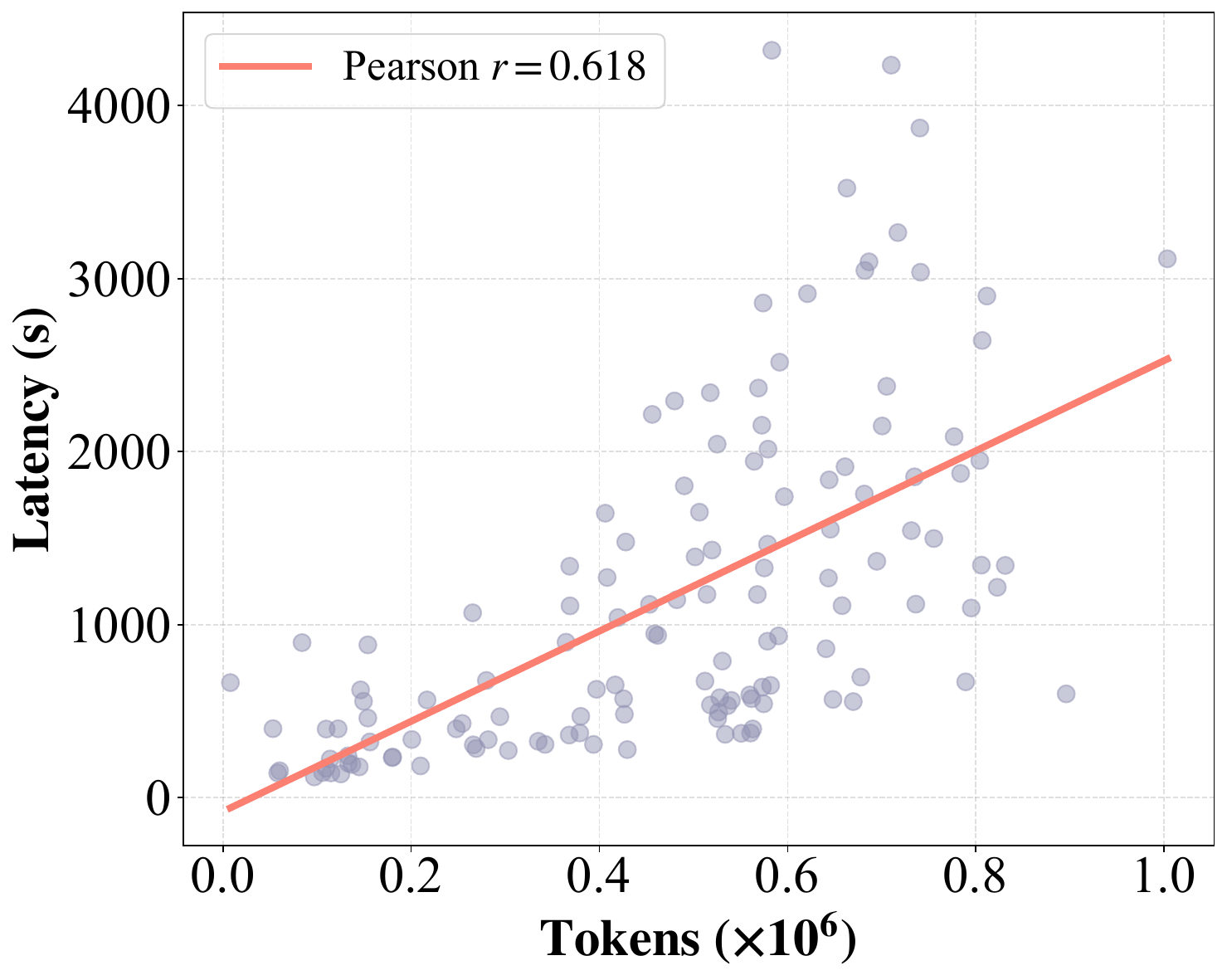}
    \caption{Correlation between token consumption and model latency.}
    \label{fig:correlation}
\end{figure}

\vpara{Distributions of Tokens, Latency, and Pass Rates.} 
Figure~\ref{fig:distribution} shows the distributions of token consumption, model latency, and pass rates across the evaluated repositories. Token consumption exhibits an approximately normal distribution. Model latency displays a strong right-skewed distribution. The pass rate distribution is distinctly bimodal, with most repositories achieving either 0\% or 100\% pass rates. 
Figure~\ref{fig:correlation} shows the correlation between token consumption and model latency. The Pearson correlation coefficient $r=0.618$ demonstrates a significant positive correlation between the two variables, indicating that higher token usage generally leads to longer processing delays.

\begin{figure*}[!ht]
    \centering
    \includegraphics[width=0.92\textwidth]{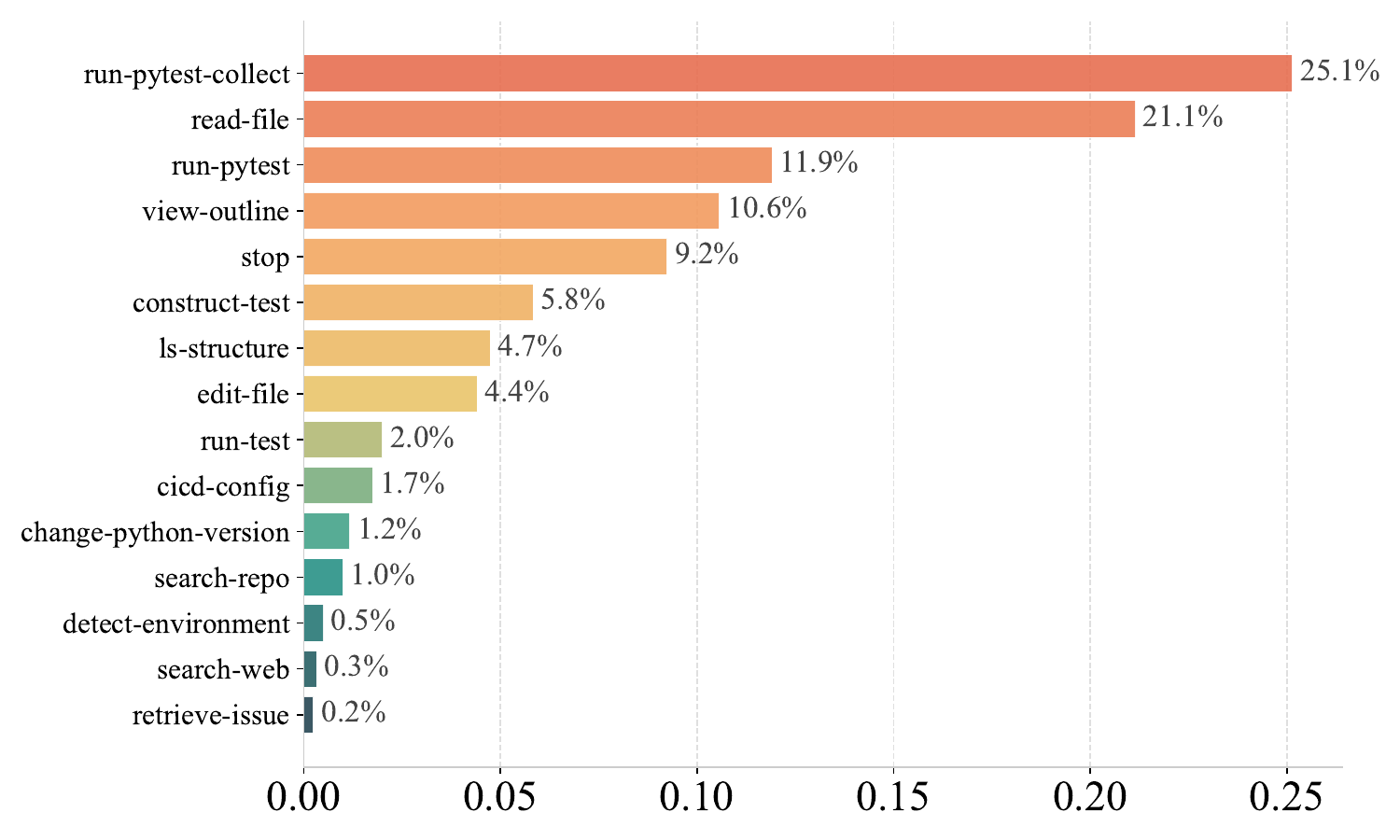}
    \caption{Tool calls distribution of \ours across Python repositories in \oursbench.}
    \label{fig:tool_distribution}
\end{figure*}

\vpara{Tool Distribution.} As shown in Figure~\ref{fig:tool_distribution}, \ours frequently invokes \texttt{run-pytest-collect} to construct test programs for determining task completion. File-related tools such as \texttt{read-file}, \texttt{view-outline}, and \texttt{ls-structure} are also commonly used. In contrast, tools associated with error recovery, such as \texttt{search-web}, \texttt{retrieve-issue}, and \texttt{change-python-version}, are invoked less often, since most issues can be resolved using the LLM’s intrinsic reasoning and debugging capabilities. Overall, the effective utilization of these tools demonstrates the soundness and rationality of our tool design.

\vpara{Action Call Illustration.} The Action Call Example for repository \texttt{abrignoni/aleapp} as shown in Table~\ref{tab:command_stats}. It begins with systematic environment awareness by inspecting the repository structure and configuration files, then adapts dynamically (e.g., switching Python versions) to resolve compatibility issues. The trajectory illustrates how \ours uses repository inspection, dependency installation, version adjustment, and validation within a 30-step budget.

\begin{table*}[t]
\centering
\renewcommand{\arraystretch}{1.2}
\caption{Command Execution Statistics Table.}
\label{tab:command_stats}
\begin{tabular}{c p{0.70\textwidth} c}
    \toprule
    \textbf{No.} & \textbf{Action (Command/Tool)} & \textbf{Duration (s)} \\ \midrule
1 & ls-structure --repo /repo --depth 3 & 1.76 \\
2 & read-file /repo/README.md & 2.07 \\
3 & read-file /repo/requirements.txt & 2.02 \\
4 & construct-test --repo /repo & 22.97 \\
5 & pip install -q pytest openai -i https://mirrors.aliyun... & 1.96 \\
6 & pip install -q -r /repo/requirements.txt -i https://mi... & 16.54 \\
7 & run-pytest-collect & 1.68 \\
8 & cd /repo & 1.68 \\
9 & python aleapp.py --help & 1.78 \\
10 & python --version & 1.69 \\
11 & change-python-version 3.10 & 12.77 \\
12 & pip install -q pytest openai -i https://mirrors.aliyun... & 2.10 \\
13 & pip install -q -r /repo/requirements.txt -i https://mi... & 11.10 \\
14 & pip install -q packaging$\geq$22 -i https://mirrors.aliyun... & 2.00 \\
15 & cd /repo & 1.68 \\
16 & python aleapp.py --help & 2.48 \\
17 & run-pytest-collect & 1.64 \\
18 & run-pytest & 1.49 \\
19 & find /repo -type f -name \texttt{"}*test*.py\texttt{"} -o -name \texttt{"}*spec*.... & 1.69 \\
20 & cd /repo & 1.69 \\
21 & python aleapp.py -p & 2.13 \\
22 & cd /repo & 1.70 \\
23 & python -c \texttt{"}import tkinter; print(\texttt{'}tkinter available\texttt{'})\texttt{"}... & 1.76 \\
24 & cd /repo & 1.70 \\
25 & python -c \texttt{"}import aleappGUI; print(\texttt{'}aleappGUI imports ... & 1.91 \\
26 & detect-environment --format text & 2.50 \\
27 & stop & 0.00 \\
\bottomrule
\end{tabular}
\end{table*}

\vpara{Cost for Each Repository.}
We report the per-repository cost of running \ours with DeepSeek-V3 on a Python environment setup.
Under the standard budget of at most 30 turns per repository, the agent achieves 63.2\% ESSR with an average cost of \$0.30 per repository.
In practical deployments, response caching further reduces repeated-token usage across turns, making the effective cost even lower.
Overall, this cost level is within an acceptable range for large-scale benchmark construction and routine use.

\vpara{Correlation between Complexity and Performance.}
Specifically, we here define project depth as the maximum depth of the repository structure tree, computed via a BFS traversal over the directory hierarchy. Based on this definition, our empirical data shows a statistically significant negative correlation between project depth and the ESSR of \ours (Pearson $r = -0.2859, p = 0.0009$). As project depth increases, the system encounters more challenges in environment configuration and dependency resolution. And the system maintains high reliability within a depth of 1-6. Beyond this range, the variance in performance increases. And it is worth noting that "depth" is not the sole bottleneck. We have observed projects with a depth of 10+ achieving a 100\% pass rate. This suggests that while depth adds complexity, the system remains capable of handling deep structures. 

\vpara{Validation of Constructed Tests (Smoke Tests).}  For repositories without any existing tests, the generated tests provide a first layer of automated validation compared to prior settings with no verification mechanism at all. Specifically, construct\_tests performs a series of checks, including validating directory structure, extracting executable commands from documentation, and identifying entry points of the program. If no explicit entry point is found, it falls back to library version checks, repository structure validation, and core module import tests. These collectively serve as reasonable surrogate tests in the absence of ground-truth test suites.

To ensure that the constructed tests are not “trivially passing,” we conducted controlled experiments by intentionally removing test files from repositories. In 70\% of the cases, our method identified valid entry points, indicating that the constructed tests provide a useful basic executability check rather than a substitute for full functional validation.

\vpara{Validation of \oursbench Design.} \oursbench primarily uses a stratified sampling strategy for sampling. And the primary goal of our two-dimensional stratified sampling strategy (Project Size × Popularity) is to ensure that RATBench accurately reflects the heterogeneity of real-world software, rather than being biased toward "trivial" cases. We conducted an additional analysis on our two-dimensional stratified sampling strategy (Project Size × Popularity), focusing on how it affects the difficulty distribution of the benchmark. Table~\ref{tab:project_size_pass_rate} shows a clear ESSR performance difference across project sizes.

\begin{table}[H]
\centering
\caption{Pass rate by project size.}
\label{tab:project_size_pass_rate}
\begin{tabular}{lc}
\toprule
\textbf{Project Size} & \textbf{Pass Rate} \\
\midrule
Small  & 65.4\% \\
Medium & 62.4\% \\
Large  & 4.8\% \\
\bottomrule
\end{tabular}
\end{table}

This demonstrates that larger projects are substantially more challenging, confirming that project size is a critical factor controlling task difficulty. Similarly, Table~\ref{tab:star_range_pass_rate} summarizes the trend across repository popularity.

\begin{table}[H]
\centering
\caption{Pass rate by repository popularity.}
\label{tab:star_range_pass_rate}
\begin{tabular}{lc}
\toprule
\textbf{Star Range} & \textbf{Pass Rate} \\
\midrule
(10, 100]   & 69.2\% \\
(100, 1000] & 61.9\% \\
(1000, +)   & 59.4\% \\
\bottomrule
\end{tabular}
\end{table}

We observe a consistent decrease in success rate as repository popularity increases, suggesting that widely-used repositories tend to exhibit higher complexity (e.g., stricter dependencies, more intricate configurations). Our stratified sampling explicitly ensures the inclusion of harder scenario (e.g., large-scale repositories), thereby preventing this bias. In conclusion, In contrast, random sampling from GitHub tends to over-represent small and less complex repositories (e.g., lightweight utilities or toy projects). As shown in Tables~\ref{tab:project_size_pass_rate} and~\ref{tab:star_range_pass_rate}, such repositories are associated with higher pass rates in our evaluation, which may lead to an overly optimistic estimate of agent capability.

\vpara{Comparison with General-purpose Code Agent.} In our current evaluation, we included SWE-agent as a representative general-purpose SE agent. We found that RAT outperforms SWE-agent by an average of 47.7\% in ESSR. This gap highlights that general agents often lack the specialized reasoning, pipelines, and toolsets (e.g., semantic image retrieval) required for environment configuration. Additionally, agents like OpenHands focus on high-level task solving. RAT, on the other hand, addresses the foundational challenge of making the repository executable. We have also conducted additional tests on a subset of 30 Python repositories, where RAT achieved an ESSR of 76.7\%, compared to 33.2\% for Claude Code, further validating the necessity of specialized configuration logic. 

\vpara{Cross-Benchmark Evaluation on Repo2Run.}
To demonstrate the generalizability of RAT and eliminate any potential evaluation bias inherent to custom datasets, we further evaluated our agent on Repo2Run, a third-party benchmark~\cite{hu2025repo2run}. Under the identical, rigorous Executable Success Rate (ESSR) metric, RAT achieves a robust ESSR of 36\%. This performance underscores our agent's strong capacity to handle complex runtime dependencies and diverse repository structures beyond \oursbench benchmark.

\vpara{Human Comparison across Programming Languages.} A broader evaluation across not only repository scales but also programming languages to provide stronger evidence for RAT’s practical applicability. We further extended the human comparison to include additional 10 repositories across diverse programming languages, covering Python, Java, Rust, and JS/TS. The expanded evaluation covers repositories with different dependency structures and configuration difficulties. The updated results are summarized in Table~\ref{tab:broader_human}

\begin{table}[H]
\centering
\caption{Experimental results across different programming languages.}
\label{tab:broader_human}
\renewcommand{\arraystretch}{1.1} 
\setlength{\tabcolsep}{6pt}
\resizebox{\columnwidth}{!}{
\begin{tabular}{l|cc|cc}
\toprule
Language & Human ESSR(\%) & RAT ESSR(\%) & Human Time(min) & RAT Time(min) \\
\midrule
Python   & 82.28          & \textbf{86.52}  & \textbf{12.07}           & 25.41 \\
Java     & 20.04 & \textbf{40.0} & \textbf{6.77}   & 18.44 \\
Rust     & \textbf{80.06}          & 60.0          & 15.13  & \textbf{5.23} \\
JS/TS    & 70.0         & 80.0         & 5.02  & \textbf{4.87}  \\
\bottomrule
\end{tabular}
}
\end{table}
These additional experiments provide preliminary evidence for RAT's behavior beyond Python-only settings.

\section{Discussion}\label{app:discussion}

\vpara{Novelty Clarification.} While prior code agents (e.g., Repo2Run, SWE-agent) utilize environment feedback, the mere adoption of a shared paradigm does not diminish our novelty. In practice, the orchestration of the workflow and tool integration contributes substantially to configuration success, as evidenced by the substantial performance gaps demonstrated in Table~\ref{tab:main-results-single-column}.

From an application perspective, unlike Repo2Run, our objective is not to introduce a new interaction paradigm, but to systematically address the open and underexplored problem of real-world environment configuration. Existing approaches, including Repo2Run, remain limited in handling heterogeneous, multi-language, and dependency-intensive settings. Our contributions lie in enabling this paradigm to operate robustly under realistic, heterogeneous, and multi-language scenarios, as well as introducing an execution-driven benchmark (RATBench) that reflects real-world distributions. These are not merely engineering refinements, but essential steps toward bridging the gap between controlled experimental setups and real-world repositories.

\vpara{Metric Validity and Rationale.}Successful build does not strictly guarantee runtime correctness, and may miss issues that only surface during execution (e.g., missing environment variables). However, for Java, Rust, Go and JS/TS languages, we adopt build success as a proxy since these languages typically provide deterministic build targets (e.g., mvn install, cargo build), but often lack standardized and unified runtime entry points or test interfaces. Under this constraint, build success is not a weak metric. It systematically verifies the integrity of static dependency resolution and effectively captures the majority of common issues, such as version mismatches and missing symbols. We recognize that this indicator does not fully capture runtime-level exceptions (e.g., missing dynamic environment variables), and we consider the integration of localized execution-based validation—such as automated entry-point probing—as a promising trajectory for future extensions.

\vpara{Discussion on the Effectiveness of Image Initialization.}
 In practice, the effectiveness of image initialization in RAT is influenced by two factors: (1) the success rate of selecting an appropriate base image, and (2) the additional cost incurred when the initial image selection is suboptimal.

For the first aspect, in practice, our semantic retrieval module achieves a high accuracy in selecting appropriate base images, which mitigates this concern in most cases. We conducted a small-scale analysis by sampling 150 repositories from RATBench and examining whether the initially selected image required subsequent modification. We observed a success rate of 87.39\%, indicating that correct initialization is achieved in the majority of cases.

For the second aspect, more importantly, RAT is explicitly designed to remain robust even when the initial choice is suboptimal. In particular, RAT includes dedicated recovery mechanisms (e.g., the change version tool) that enable iterative environment adjustment and resolution of incompatibilities caused by incorrect initialization.

\vpara{Principles of Tool Design.}
The design of tools follows three key principles. (1) Abstraction of high-level actions: Compared to raw bash commands, tools encapsulate recurring and structured operations (e.g., dependency handling), which are otherwise difficult for LLMs to reliably compose through low-level command sequences. (2) Context efficiency: Tools help manage the agent’s interaction context by avoiding verbose command outputs, thereby reducing token consumption and improving stability. (3) Search space reduction: Environment configuration is inherently a long-horizon search problem. Direct shell interaction leads to a combinatorial explosion of possible command sequences, making exploration inefficient and error-prone. Tool abstractions constrain the action space into structured operations, significantly improving both efficiency and robustness.

Regarding the number of tools, we do not include them arbitrarily. Instead, we perform empirical filtering and retain only frequently used tools, while removing those rarely invoked by the agent (as shown in Figure~\ref{fig:tool_distribution} in the Appendix). We also agree that an excessive number of tools can negatively affect performance, and our design reflects this consideration.

\vpara{Impact of Data Leakage.} We claim that data leakage has a limited impact on \ours: (1) RATBench requires dynamic environment interaction (e.g., installation, debugging), making it execution-dependent rather than memorization-based; (2) it includes diverse, long-tail repositories, which reduces the likelihood of memorization; (3) large performance gaps across agents using the same LLM (i.e., 15.5\%) suggest that memorization alone is insufficient to explain the results. Furthermore, to minimize potential leakage, we will continue to mitigate it by incorporating unseen repositories.

\vpara{Independence Verification of RAT and RATBench.} The design of RAT and the construction of RATBench are independent. RATBench is collected through unbiased stratified sampling, and the repositories contain general information available to environment configuration agents, such as repository structure, dependency manifests, and build/test artifacts, rather than RAT-specific signals, tools, or verification routines. All compared methods receive the same repository inputs and are evaluated under identical verification protocols without any additional information introduced for RAT.

Furthermore, to address this concern, we evaluate RAT on an external repository environment configuration benchmark, Repo2Run, which can be deem constructed independently from RATBench. Without access to any RATBench-specific information or procedures, RAT still consistently outperforms previous approaches, achieving an improvement of 20.5\% over the strongest baseline. This demonstrates that RAT’s gains come from its general environment reasoning and tool-augmented configuration framework rather than benchmark-specific optimization.

\section{An Example of Extensibility - Adding Go language}\label{app:example}
In this section, we discuss how to manually adapt the agent to a new programming language. Our framework is designed for lightweight extensibility, so supporting a new language (e.g., C++ or Go) requires only minimal, modular additions. Specifically, this involves adding simple language-specific heuristics (e.g., detecting CMakeLists.txt or Makefile), defining the corresponding build/test commands (e.g., cmake, make), and extending the language detector with a few additional rules.

Additionally, a basic container template (e.g., with GCC/Clang and CMake) and optional dataset entries can be incorporated following the existing pipeline. Importantly, these changes do not require modifying the core framework, as RAT already encapsulates language-specific logic within a unified abstraction.

\begin{figure}[H]
    \centering
    \includegraphics[width=\linewidth]{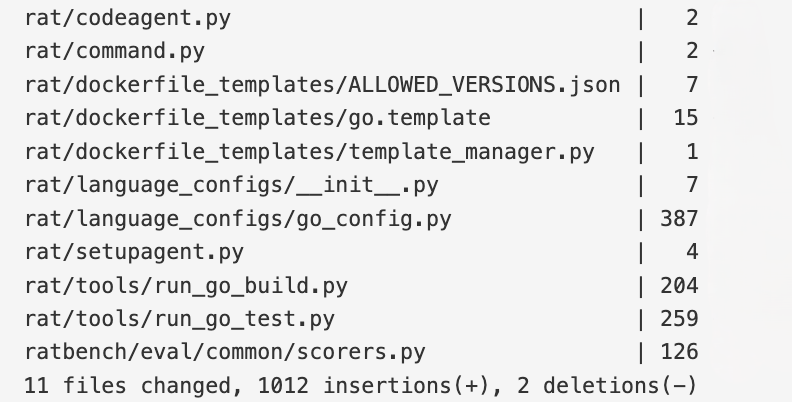}
    \caption{Code changes required to extend \ours with Go support.}
    \label{fig:go_diff}
\end{figure}

Regarding manual intervention, we here provide a detailed quantitative analysis of the manual effort required to extend RAT to support a new language. Taking the addition of Go support as an example, the total code changes amount to only 1,014 lines, and RAT with DeepSeek-V3 achieves 72.2\% ESSR, with the code changes summarized in Figure~\ref{fig:go_diff}.

We further break down this effort as follows:

1. The majority of the additional code is dedicated to RATBench evaluation (589 lines, 58.1\%), which requires language-specific test runners (run\_go\_build, run\_go\_test) and corresponding scorers.
2. The remainder extends the RAT agent itself (425 lines, 41.9\%), consisting of: language-specific plan templates (155 lines, 15.3\% — this portion does involve language-specific patterns, but can be replaced by the auto plan mode), environment detection and version management tools (15 lines, ~0.01\%, e.g., go version), and miscellaneous boilerplate code for framework compatibility (255 lines, 25.1\%).

As the breakdown above demonstrates, these heuristics are designed following a strict minimalist principle to avoid introducing excessive human priors. The core design and advantage of RAT lies in its workflow architecture rather than these language-specific heuristics. Therefore, this case study suggests that adding support for a new language requires limited changes to the RAT framework, although broader evidence across more languages is needed to fully characterize extensibility.

\begin{figure}[!h]
    \centering
    \includegraphics[width=0.5\textwidth]{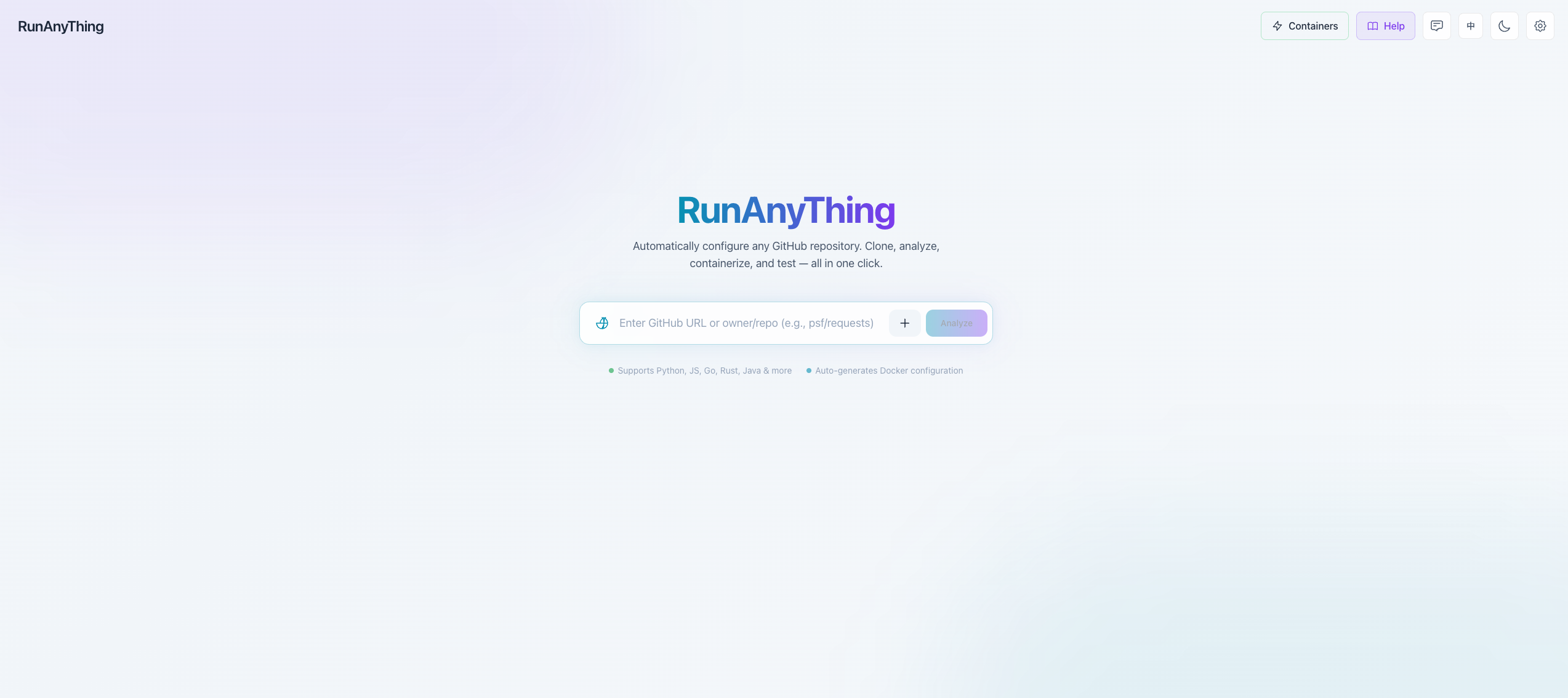}
    \caption{Web Interface for \ours}\label{fig:cover}
\end{figure}

\section{Web Interface}\label{app:app}

We also provides web interfaces for ease of use and developed \ours (RunAnyThing) as shown in Figure~\ref{fig:cover}. Built upon LLM Agents and Docker, the system integrates the entire pipeline, \emph{repository parsing, environment construction, dependency installation, and test verification}, into an end-to-end workflow. Users can provide either a GitHub repository URL or upload a ZIP archive, after which the system automatically identifies the programming language and dependency management framework, recommends and constructs an appropriate base image, executes environment configuration and testing inside containers, and finally outputs reusable Dockerfiles/container images together with structured results (e.g., test status, key commands, and execution logs).

For frontend and backend, \ours adopts a full-stack architecture based on FastAPI and React/TypeScript, and provides real-time progress tracking across five stages (cloning, analysis, building, configuration, and testing) through Server-Sent Events (SSE). The system also offers an interactive container terminal, repository static analysis capabilities (including code metrics, dependency coupling analysis, and key function identification), bilingual Chinese–English interfaces, and a feedback analytics module, supporting a closed-loop workflow from automated execution to human evaluation. Overall, the application significantly reduces the cost of configuring complex project environments while improving experiment reproducibility and replication efficiency.

\section{Broader Impacts}\label{app:impact}
\ours is expected to play an important role in multiple areas in the future, with broader impacts:

(1) \emph{Scalable Data Synthesis}: By automatically transforming static repositories into verifiable, runnable datasets, the Agent enables large-scale benchmark construction and generates precise execution traces, supporting post-training LLM development. (2) Execution-based Reinforcement: Automated setup allows agents to leverage true functional feedback, moving reward models beyond static or heuristic signals toward execution-aware learning. (3) Executable Analysis: \ours facilitates systematic analysis of code executability.

Moreover, in our future plans, given a functional requirement (e.g., ``implement real-time translation''), \ours can bypass manual coding by identifying stable repositories, automatically configuring environments, resolving dependencies, extracting essential logic, and integrating across projects. It continuously enables one-click deployment of fully functional, ready-to-run environments. This end-to-end automation accelerates experimentation, ensures reproducible execution, and supports scalable evaluation in both interactive and headless SaaS settings.

\tcbset{
appendixpromptbox/.style={
    enhanced jigsaw,
    breakable,
    colback=blue!2!white,
    colframe=blue!45!black,
    colbacktitle=blue!45!black,
    coltitle=white,
    fonttitle=\bfseries,
    boxrule=0.4pt,
    arc=1mm,
    left=1mm,
    right=1mm,
    top=1mm,
    bottom=1mm,
    before skip=0.5\baselineskip,
    after skip=0.5\baselineskip
}
}

\clearpage
\onecolumn
\raggedbottom
\section{Prompts}
\label{app:prompts}
\begin{tcolorbox}[appendixpromptbox, title={Dockerfile Template for Pipreqs.}]
\begin{verbatim}
FROM python:3.10
WORKDIR /
RUN pip install pytest pytest-xdist && pip install pipdeptree

COPY . /repo
WORKDIR /repo
# Ensure we use requirements.txt generated by pipreqs
RUN if [ -f requirements_pipreqs.txt ]; then pip install -r requirements_pipreqs.txt; 
fi

# Try pytest collect to sanity-check the environment
RUN pytest --collect-only -q || true
\end{verbatim}
\end{tcolorbox}

\begin{tcolorbox}[appendixpromptbox, title={Prompt for Zero-Shot.}]
\begin{verbatim}
You are an expert DevOps engineer specializing in containerization.
Your task is to generate a Dockerfile that can successfully setup a given
repository.

Requirements:
1. Use an appropriate base image for the programming language
2. Set up the correct working directory
3. Copy all necessary files
4. Install all dependencies
5. Set appropriate environment variables if needed

Output ONLY the Dockerfile content without any explanation or markdown code
blocks.

Based on the following repository information, generate a production-ready
Dockerfile:

Repository: {repo_name}

{repo_context}

Generate a Dockerfile that:
- Installs all required dependencies
- Sets up all variables and configuration correctly
- Uses best practices for the detected language/framework

Output the Dockerfile content directly:
\end{verbatim}
\end{tcolorbox}

\begin{tcolorbox}[appendixpromptbox, title={Configuration for SWE-agent.}]
\begin{verbatim}
agent:
  model:
    name: deepseek/deepseek-chat
    per_instance_call_limit: 30
  templates:
    system_template: |-
      You are an expert DevOps engineer specialized in repository 
      environment setup and configuration.
      ...
    instance_template: |-
      <uploaded_files>
      {{working_dir}}
      </uploaded_files>
      I have uploaded a Python repository in {{working_dir}}.

      <setup_requirements>
      {{problem_statement}}
      </setup_requirements>

      Your task is to configure this repository so it can run and pass all 
      tests.

      ## Workflow
      1. Explore Repository: ...
      2. Analyze Dependencies: ...
      3. Install Dependencies: ...
      4. Verify Installation: ...
      5. Fix Environment Issues: ...
      6. Validate Setup: ...

      ## Code Modification Policy
      ...

      ## Important Notes
      - Focus on installing dependencies and configuring the environment
      - ...

    problem_statement_template: |-
      This is a {{language}} repository that requires environment setup and
      configuration.

      ## Objective
      Configure the repository to be ready for running tests successfully.

      ## Detailed Steps

      ### 1. Repository Analysis
      ...

      ### 2. Dependency Installation
      ...

      ### 3. Environment Configuration
      ...

      ### 4. Installation Verification
      ...

      ### 5. Troubleshooting
      ...

      ## Success Criteria
      The repository is considered properly configured when:
      - All dependencies are installed successfully
      - The environment is properly set up (config files, env vars, etc.)
      - ...

      ## Constraints
      ...
    next_step_template: |-
      Observation:
      {{observation}}
    next_step_no_output_template: |-
      Your command ran successfully, but produced no output.
  tools:
    env_variables:
      PAGER: cat
      MANPAGER: cat
      LESS: -R
      PIP_PROGRESS_BAR: 'off'
      TQDM_DISABLE: '1'
      GIT_PAGER: cat
    bundles:
      - path: tools/registry
      - path: tools/edit_anthropic
      - path: tools/review_on_submit_m
    registry_variables:
      USE_FILEMAP: 'true'
      SUBMIT_REVIEW_MESSAGES:
        - |
          ## Pre-Submission Checklist
          ...

          ## Review Your Changes
          ...

          ## Action Required
          ...
    enable_bash_tool: true
    parse_function:
      type: function_calling
  history_processors:
    - type: cache_control
      last_n_messages: 2
env:
  deployment:
    type: docker
    image: python:3.10-slim
    remove_container: false  # Do not auto-remove container
    remove_images: false     # Do not remove images
  repo:
    type: local
    path: ./repo
\end{verbatim}
\end{tcolorbox}

\begin{tcolorbox}[appendixpromptbox, title={Prompt for ImageRetriever in Initialization.}]
\begin{verbatim}
You are a Docker image selection expert. Infer the best Docker base image
from the information below.

## Allowed versions (ALLOWED_VERSIONS.json)
```json
{json.dumps(allowed_versions, ensure_ascii=False, indent=2)}
```

## Repository config files
{config_str if config_str else "(No config files found)"}

## README
{readme_content if readme_content else "(No README found)"}

## Requirements
1. **Identify the primary language** (...)
2. **Analyze version requirements**: ...
3. **Select the best image**:
   - Prefer selecting language/version/variant from ALLOWED_VERSIONS.json
   - If the project requires a specific version, pick the closest allowed
   version
   - If version is unclear, use default_version
   - For variant (e.g. slim, alpine), use default_variant by default
4. **Extract dependency info** (optional): key frameworks and dependencies

## Output format (raw JSON only; no Markdown)
{{
  "language":  ...,
  "version": ...,
  "variant": ...,
  "base_image": ...,
  "full_image": ...,
  "reason": ...,
  "confidence": ...,
  "frameworks": ...,
  "dependencies": ...
}}

## Important rules
1. **Version constraint**: ...
2. **Variant constraint**: ...
3. ...

## Example output
{{
  ...
}}
\end{verbatim}
\end{tcolorbox}

\begin{tcolorbox}[appendixpromptbox, title={Prompt for Standard Plan Mode.}]
\begin{verbatim}
You are an expert in environment setup. You may refer to files and structures 
in the repository such as requirements.txt, setup.py, etc., and use 
dependency inference tools like pipreqs to install third-party libraries
inside the specified Docker image. This ensures the repository can be set up 
successfully and the specified tests can run.

Note: This repository originally has no Dockerfile, or an existing Dockerfile 
has been removed. Do not attempt to use the repository's original Dockerfile
information.

Workflow:
0. Explore the repository to understand its full structure and the image 
environment.
1. Find the program entry point; create a usable test case; check whether 
tests pass without extra setup.
2. Read key documentation, including ...
3. Inspect repository directories and read environment-related files ...
4. Collect dependency lists: find dependency files in the repo root ...
5. Install dependencies based on collected files...
6. Check whether tests pass; if so, call stop.

CLI tool instructions:
All operations run inside Docker container {image_name}
Think about what to do next, then wrap commands with ...

Note: Do not make large changes to /repo; only necessary adjustments.
Keep commands on a single line when possible using &&. Avoid multi-line 
commands, backslash continuations, and HERE-DOCs (<<).

Available tools (callable tools, not built-in terminal commands):
{tools_list}

Important: ...

Special note: ...
\end{verbatim}
\end{tcolorbox}

\begin{tcolorbox}[appendixpromptbox, title={Prompt for Automated Plan Mode.}]
\begin{verbatim}
You are an expert in environment setup. You must configure the environment
according to a custom plan file `plan.md`. You may use `/repo/plan.md` as
your task plan (it does not exist initially). This file contains the
environment setup goal, phases, and progress.
The primary language of the repository is {language}, so configure the 
environment accordingly.

## Two-phase workflow
### Phase 1: Check and create the plan (if missing)
1. First check whether `/repo/plan.md` exists.
2. If it does not exist, you should:
   - Explore repository structure and understand the project
   - Create an initial `plan.md` based on your analysis (make steps concrete;
avoid too many generic steps)
   - Use the `edit-file` tool to create the file; use the template below:

```markdown
# Task Plan: Environment Configuration for {image_name}

## Goal
[Goal of environment setup, e.g. configure {image_name} environment so tests
pass. You can execute at most {max_turn} steps, so keep the plan simple.]

### Phase 1: Repository Analysis
<!-- Analyze repository structure and dependencies -->
- [ ] Explore repository structure
- [ ] Identify main entry point
- [ ] Read README and documentation
- [ ] Find dependency files (requirements.txt, pyproject.toml, etc.)
- **Status**: pending

### Phase 2: Dependency Installation
<!-- Install dependencies -->
- [ ] Install system dependencies (if any)
- [ ] Handle version conflicts (if any)
- [ ] Install development dependencies 
- **Status**: pending

### Phase 3: Environment Configuration
<!-- Environment configuration -->
- [ ] Configure environment variables
- [ ] Set up database (if needed)
- [ ] Set up paths and permissions
- **Status**: pending

### Phase 4: Testing & Validation
<!-- Testing and validation -->
- [ ] Run basic tests
- **Status**: pending

## Current Phase
Phase 1

## Notes
[Any important notes]
```
3. If the file already exists, read and understand the existing plan.

### Phase 2: Execute the plan
1. Read the current plan and understand the current phase and goal.
2. Find the current phase (an unchecked `[ ]` phase).
3. Execute tasks in the current phase:
   - If the phase status is not marked, update it to `in_progress` using
`edit-file`
   - Execute tasks (explore/install/configure/test)
   - After finishing the phase, update `[ ]` to `[x]` and set status to
`complete`
4. Continue to the next phase until all phases are complete

## Important rules
...

## Response format requirements
...

**Example:**
...

## Initial action guide
...

Available tools:
{tools_list}
\end{verbatim}
\end{tcolorbox}

\begin{tcolorbox}[appendixpromptbox, title={Prompt for Calling Different Actions.}]
\begin{verbatim}
You are an expert proficient in testing. The current environment has been
built using the repository's Dockerfile (image: {image_name}), and the 
environment configuration should be complete.

Your main tasks are:
1. Verify that the environment is working correctly
2. Create test cases
3. Run tests and ensure they pass

Workflow:
1. Quick environment verification: Check if key commands and dependencies 
are available
2. Use the construct-test tool to create test cases
3. Run tests (run-pytest-collect and run-pytest)
4. ...

CLI Tool Usage Instructions:
All operations are performed inside Docker container {image_name}
... for example:

### Thought: ...
### Action:
{BASH_FENCE[0]}
...
{BASH_FENCE[1]}

Available tools (callable but not terminal built-in commands):
{tools_list}

Important Notes:
1. Environment has been built via repository Dockerfile, most dependencies 
should already be installed
2. ...

Special Note:
...
\end{verbatim}
\end{tcolorbox}

\begin{tcolorbox}[appendixpromptbox, title={Prompt for Go Configuration.}]
\begin{verbatim}
you are an expert proficient in go environment configuration.
your workflow is:
0. explore the repository to understand the complete project structure and the 
image configuration.
1. read and understand important documentation in the repository, which may 
include but is not limited to readme.md, contributing.md, *.md, *.txt, content 
in docs/, etc.
2. read the repository directory and files related to environment configuration, 
such as go.mod, go.sum, .go-version, etc. consider other files and structures 
that may be used for environment configuration.
3. collect dependency information:
   - check dependencies in go.mod file
   - check if it's a multi-module project (multiple go.mod files)
   - check required go version
4. build the project:
   - use the run-go-build tool to build the project
   - if build fails, analyze errors and fix
5. run build to verify configuration. if build succeeds, call stop.
...
\end{verbatim}
\end{tcolorbox}

\end{document}